\pdfoutput=1
\documentclass[11pt]{article}

\usepackage{amsmath,amssymb}
\usepackage{amscd,amsthm}
\usepackage{amscd}

\usepackage{multirow}
\usepackage{accents}
\usepackage{hyperref}
\usepackage{color}
\usepackage[usenames,dvipsnames]{xcolor}
\usepackage{graphicx}
\usepackage{enumitem}

\usepackage{algorithm}
\usepackage{algorithmic}

\usepackage{setspace}

\usepackage{graphicx} 

\usepackage[
pass,
]{geometry}

\newgeometry{top=1in, bottom=1in, left=1in, right=1in}

\newfont{\mycrnotice}{ptmr8t at 7pt}
\newfont{\myconfname}{ptmri8t at 7pt}

\newtheorem{theorem}{Theorem}
\newtheorem{lemma}{Lemma}

\newtheorem{corollary}{Corollary}
\newtheorem{proposition}{Proposition}

\newtheorem{definition}{Definition}

\newcommand{\A}{\mathcal{A}}
\newcommand{\T}{\mathfrak{T}}
\newcommand{\RR}{\mathfrak{R}}
\newcommand{\LL}{\mathfrak{L}}

\newcommand{\PP}{\wp}

\newcommand{\astr}{\mathbf{a}}

\newcommand{\Opt}{\mathrm{Opt}}

\newcommand{\Reg}{\mathrm{Reg}}

\newcommand{\SReg}{\mathrm{SReg}}
\newcommand{\Sur}{\mathrm{Sur}}

\newcommand{\n}{\mathfrak{n}}
\newcommand{\m}{\mathfrak{m}}
\newcommand{\rt}{\mathfrak{r}}
\newcommand{\lf}{\mathfrak{l}}

\newcommand{\Ind}{\mathbb{I}}
\newcommand{\fa}{\:\forall}

\newcommand{\g}{\gamma}
\newcommand{\ve}{\varepsilon}
\newcommand{\e}{\epsilon}

\newcommand{\gb}{{\boldsymbol\g}}

\newcommand{\argmax}{\mathop{\mathrm{argmax}}}

\newcommand{\ASet}{\mathbf{A}}
\newcommand{\pre}{\mathtt{pre}} 
\newcommand{\treeroot}{\mathfrak{e}}

\title{On consistency of optimal pricing algorithms in repeated posted-price auctions with strategic buyer}
\date{July 11, 2017}

\author{
	{\bf Alexey Drutsa}\\ Yandex\thanks{16, Leo Tolstoy St., Moscow, Russia, 119021 (www.yandex.com)}, MSU\thanks{Lomonosov Moscow State University, Faculty of Mechanics and Mathematics; GSP-1, 1 Leninskiye Gory, Main Building, Moscow, Russia, 119991} \\ {\tt\small adrutsa@yandex.ru}
}

\begin{document}

\maketitle

\begin{abstract}
We study revenue optimization learning algorithms for repeated posted-price auctions where a seller interacts with a single strategic buyer that holds a fixed private valuation for a good and seeks to maximize his cumulative discounted surplus.
For this setting, first, we propose a novel algorithm that never decreases offered prices and has  a tight strategic regret bound in $\Theta(\log\log T)$ under some mild assumptions on the buyer surplus discounting.
This result closes the open research question on the existence of a no-regret horizon-independent weakly consistent pricing.
The proposed algorithm is inspired by our observation that a double decrease of offered prices in a weakly consistent algorithm  is enough to cause a linear regret.
This motivates us to construct a novel transformation that maps a right-consistent algorithm to a weakly consistent one that never decreases offered prices.

Second, we outperform the previously known strategic regret upper bound of the algorithm PRRFES, where the improvement is achieved by means of a finer constant factor $C$ of the principal term $C\log\log T$ in this upper bound.
Finally, we generalize results on strategic regret previously  known for geometric discounting of the buyer's surplus to discounting of other types, namely: 
the optimality of the pricing PRRFES to the case of geometrically concave decreasing  discounting; and linear lower bound on the strategic regret of a wide range of horizon-independent weakly consistent algorithms to the case of arbitrary discounts.
\end{abstract}


\newpage

\section{Introduction}
\label{sec_Intro}
Revenue maximization in online advertising represents one of the most important development direction in leading Internet companies (such as 
real-time ad exchanges~\cite{2014-WWW-Gomes,2015-ManagSci-Balseiro},
search engines~\cite{2007-IJIO-Varian,2009-WWW-Aggarwal,2009-SIGIR-Zhu,2013-IJCAI-He,2016-WSDM-Charles}, social networks~\cite{2014-KDD-Agarwal}, etc.), where a large part of advertisement inventory is sold via widely applicable second price auctions~\cite{2013-IJCAI-He,2014-ICML-Mohri}, including their generalizations as GSP~\cite{2007-IJIO-Varian,2012-WWW-Lucier,2014-ECRA-Sun,2016-WSDM-Charles} and Vickrey-Clarke-Groves (VCG)~\cite{2009-AER-Varian,2014-AER-Varian} auctions.
Optimal revenue here is mostly controlled by means of  reserve prices, whose proper setting is studied both by game-theoretical methods~\cite{1981-MOR-Myerson,2009-Book-Krishna} and by machine learning approaches~\cite{2007-Book-Nisan,2013-SODA-Cesa-Bianchi,2013-IJCAI-He,2013-NIPS-Amin,2014-WWW-Hummel,2014-ICML-Mohri,2014-KDD-Yuan,2014-ECRA-Sun,2014-NIPS-Mohri,2015-UAI-Mohri,2016-WWW-Paes,2016-WWW-Rudolph,2016-EC-Roughgarden,2017-WWW-Drutsa}. 
A large number of online auctions run, for instance, by ad exchanges involve only a single bidder~\cite{2013-NIPS-Amin,2014-NIPS-Mohri,2017-WWW-Drutsa}, and, in this case, a second-price auction with reserve is equivalent to a \emph{posted-price auction}~\cite{2003-FOCS-Kleinberg} where the seller sets a reserve price for a good (e.g., an ad
space) and the buyer decides whether to accept or reject this price (i.e., to bid above or below it).

In this work, we focus on a scenario when the seller \emph{repeatedly} interacts through a posted-price mechanism with the \emph{same} strategic buyer that holds a \emph{fixed} private valuation for a good and seeks to maximize his cumulative discounted surplus~\cite{2013-NIPS-Amin}.
At each round of this game, the seller is able to chose the price based on previous decisions of the buyer, i.e., to apply a deterministic online learning (discrete) algorithm.
The seller's goal is to maximize his cumulative revenue over a finite number of rounds~$T$ (the time horizon), which is generally reduced to \emph{regret} minimization\footnote{In our study, the regret is the difference between  the revenue that would have been earned by offering the buyer's valuation and the seller's revenue; it is optimized for the  worst-case buyer valuation (see Sec.~\ref{subsec_Setup} and in~\cite{2003-FOCS-Kleinberg,2014-NIPS-Mohri,2017-WWW-Drutsa}).}, and the seller  seeks thus for a \emph{no-regret} pricing algorithm, i.e., with a sublinear regret on $T$~\cite{2013-NIPS-Amin,2014-NIPS-Mohri,2014-NIPS-Amin,2015-NIPS-Mohri,2016-SSRN-Chen,2017-WWW-Drutsa}.

For this setting, the algorithm PRRFES with tight strategic regret bound in $\Theta(\log \log T)$ was recently proposed for the case when the buyer's cumulative surplus is geometrically discounted~\cite{2017-WWW-Drutsa}. 
This algorithm is horizon-independent and right-consistent (i.e., it never proposes prices lower than earlier accepted ones). 
However, its key peculiarity consists in its ability to decrease an offered price after its rejection, but then to revise it and, moreover, to propose higher prices than this  one in subsequent rounds (not satisfying thus the left consistency), such a behavior of the algorithm may be confusing to a buyer. 
Despite the fact that there does not exist a no-regret horizon-independent algorithm with the fully consistent property (both right, and left), the question on the existence of such algorithm with the consistent property in weak sense remains open~\cite{2017-WWW-Drutsa}.

The primary research goal of our study is, first, to find a no-regret weakly consistent pricing algorithm and to resolve thus the open research question.
Second, we are aimed to improve the currently best known upper bounds on strategic regret\footnote{Since these bounds are tight~\cite{2017-WWW-Drutsa}, we are aimed to improve the constant factor $C$ of its principal term $C\log\log T$.} and to generalize results on them to families of buyer discount sequences that are wider than geometric ones.

We propose a novel algorithm that never decreases offered prices and can be applied against strategic buyers with a tight regret bound in $\Theta(\log\log T)$ under some mild assumptions on the discounting of the buyer's surplus (Th.~\ref{th_prePRRFES_regret_upper_bound}).
This result constitutes \emph{the first contribution of our work} and closes the open research question on the existence of a no-regret horizon-independent weakly consistent pricing.
The key idea of this algorithm is based on our observation that a double decrease of offered prices by a weakly consistent algorithm  is enough to cause a linear regret (Lemma~\ref{lemma_WC_SReg_LowerBound}).
This motivates us to propose a novel transformation that being applied to a right-consistent algorithm results in a weakly consistent one which has no decrease of offered prices (Lemma~\ref{lemma_pre_RC_alg}).

\emph{The second contribution} consists in a novel strategic regret upper bound  for the algorithm PRRFES which outperforms the previously known one from~\cite{2017-WWW-Drutsa}. This is achieved through obtaining a finer expression for the constant factor $C$ of the principal term $C\log\log T$ of this upper bound that can be optimized by adjusting the algorithm's parameter  (Th.~\ref{th_PRRFES_regret_upper_bound}).
Finally, \emph{our work contributes also}
the generalization of the tight strategic regret bound of the pricing PRRFES to the case of geometrically concave decreasing  discounting of the buyer surplus (Th.~\ref{th_PRRFES_regret_upper_bound})
and the generalization of the previously known linear lower bound on the strategic regret of a wide range of horizon-independent weakly consistent algorithms to the case of arbitrary discounts (Lemma~\ref{lemma_WC_SReg_LowerBound} and Cor.~\ref{cor_RWC_SReg_LowerBound}).


\section{Preliminaries}
\label{sec_Fwk}
\subsection{Setup of repeated posted-price auctions}
\label{subsec_Setup}
We consider the following scenario of \emph{repeated posted-price auctions}~\cite{2013-NIPS-Amin,2014-NIPS-Mohri,2017-WWW-Drutsa}.
The seller repeatedly proposes goods (e.g., advertisement spaces) to a single buyer over $T$ rounds (the time horizon): one good per round. 
The buyer holds a \emph{fixed private valuation} $v \in [0; +\infty)$ for a good, i.e., the valuation $v$ is unknown to the seller and is equal for goods offered in all rounds. 
At each round $t\in\{1,\ldots,T\}$, a price $p_t$ is offered by the seller, and an allocation decision $a_t\in\{0,1\}$ is made by the buyer: $a_t=1$, when the buyer accepts to buy a currently offered good at that price, $0$, otherwise. 
Thus, the seller applies a \emph{(pricing) algorithm} $\A$ that sets prices $\{p_t\}_{t=1}^{T}$ in response to buyer decisions $\astr=\{a_t\}_{t=1}^{T}$ referred to as a \emph{(buyer) strategy}. We consider the deterministic online learning case when the price $p_t$ at a round $t\in\{1,\ldots,T\}$ can depend only on the buyer's actions during the previous rounds $\astr_{1:t-1}$\footnote{We use a notation for a part of a strategy $\astr_{t_1:t_2}=\{a_t\}_{t=t_1}^{t_2}$.}. Following~\cite{2017-WWW-Drutsa}, we are studying algorithms that does not depend on the horizon $T$ since it is very natural in practice (e.g., of ad exchanges) that the seller does not know in advance the number of rounds~$T$ that the buyer wants to interact with him. Let $\ASet$ be the set of such algorithms.

Hence, given an algorithm $\A\in\ASet$, a strategy $\astr$ uniquely defines  the corresponding price sequence $\{p_t\}_{t=1}^{T}$.
Hence, given a pricing algorithm $\A$, a buyer strategy $\astr = \{a_t\}_{t=1}^{T}$ uniquely defines  the corresponding price sequence $\{p_t\}_{t=1}^{T}$, which, in turn, determines  the seller's total revenue $\sum_{t=1}^T a_tp_t$.
This revenue is usually compared to the revenue that would have been earned by offering the buyer's valuation $v$ if it was known in advance to the seller~\cite{2003-FOCS-Kleinberg,2013-NIPS-Amin,2014-NIPS-Mohri,2017-WWW-Drutsa}. This leads to the definition of the \emph{regret} of the algorithm $\A$  that faced a buyer with the valuation $v\in[0,1]$ following the (buyer) strategy $\astr$ over $T$ rounds as
$
\Reg(T,\A,v,\astr):= \sum_{t=1}^T(v - a_tp_t).
$

%
Following a standard assumption in mechanism design that matches the practice in ad exchanges~\cite{2014-NIPS-Mohri}, the pricing algorithm $\A$, used by the seller, \emph{is announced to the buyer in advance}. In this case, the buyer can act strategically against this algorithm: we assume that the buyer follows the optimal strategy $\astr^{\Opt}(T,\A,v,\gb)$ that maximizes the buyer's $\gb$-discounted surplus~\cite{2013-NIPS-Amin}: 
\vspace{-0.3cm}
$$
\Sur_\gb(T,\A,v,\astr):= \sum_{t=1}^T\g_ta_t(v - p_t),
\vspace{-0.3cm}
$$
i.e., $\astr^{\Opt}(T,\A,v,\gb):= \argmax_{\astr}\Sur_\gb(T,\A,v,\astr)$, where $\gb=\{\g_t\}_{t=1}^{\infty}$ is \emph{the discount sequence}, which is assumed positive,  $\g_{t}>0 \fa t\!\in\!\mathbb{N}$, with convergent sums, $\sum_{t = 1}^{\infty} \g_t \!<\!\infty$.
Thus, we define \emph{the strategic regret} of the algorithm $\A$ that faced a strategic buyer with valuation $v\in[0,1]$ over $T$ rounds as 
$$
\SReg(T,\A,v,\gb) := \Reg\big(T,\A,v,\astr^{\Opt}(T,\A,v,\gb)\big).
$$
Hence, we consider a two-player non-zero sum repeated game with incomplete information and unlimited supply, introduced by Amin et al.~\cite{2013-NIPS-Amin} and considered in~\cite{2014-NIPS-Mohri,2017-WWW-Drutsa}: the buyer seeks to maximize his surplus, while the seller's objective is to minimize his strategic regret (i.e., maximize his revenue). 
Note that only the buyer's objective is discounted over time (not the seller's one), which is motivated by the observation that sellers are far more willing to wait for revenue than buyers are willing to wait for goods in important real-world markets like online advertising~\cite{2013-NIPS-Amin,2014-NIPS-Mohri}.

In  our setting, following~\cite{2003-FOCS-Kleinberg,2013-NIPS-Amin,2014-NIPS-Amin,2014-NIPS-Mohri,2015-NIPS-Mohri,2017-WWW-Drutsa}, we are interested  in  algorithms  that  attain $o(T)$ strategic regret  (i.e.,  the  averaged  regret  goes  to  zero  as $T\rightarrow\infty$) for  the  worst-case valuation $v\in[0,1]$, i.e., we  say that an algorithm $\A$ is \emph{no-regret} when $\sup_{v\in[0,1]}\Reg(T,\A,v,\astr^{\Opt})=o(T)$. 
Namely, we seek for algorithms that have the lowest possible strategic regret upper bound of the form $O(f(T))$ and treat their optimality in terms of $f(T)$ with the slowest growth as $T\rightarrow\infty$ (the  averaged  regret  has thus the best rate of convergence to zero).

\subsection{Notations and auxiliary definitions}
\label{subsec_Notations}
Similarly to~\cite{2017-WWW-Drutsa}, a deterministic pricing algorithm $\A$ can be associated with an infinite complete binary tree $\T(\A)$~\cite{2003-FOCS-Kleinberg,2014-NIPS-Mohri} (since we consider horizon-independent algorithms).
Each node $\n\in\T(\A)$\footnote{For simplicity, if $\n$ is a node of a tree $\T$, we write $\n\in\T$.} is labeled with the price $p^{\n}$ offered by $\A$. The right and left children of $\n$ are denoted by $\rt(\n)$ and $\lf(\n)$ respectively. The left (right) subtrees rooted at the node $\lf(\n)$ ($\rt(\n)$ resp.) are denoted by $\LL(\n)$ ($\RR(\n)$ resp.). The operators $\lf(\cdot)$ and $\rt(\cdot)$ sequentially applied $s$ times to a node $\n$ are denoted by  $\lf^s(\n)$ and $\rt^s(\n)$ respectively, $s\in\mathbb{N}$.
The root node of a tree $\T$ is denoted by $\treeroot(\T)$.

So, the algorithm's work flow is following: it starts at the root $\treeroot(\T(\A))$ of the tree $\T(\A)$ by offering the first price $p^{\treeroot(\T(\A))}$ to the buyer; at each step $t<T$, if a price $p^{\n}, \n\in\T(\A),$ is accepted, the algorithm moves to the right node $\rt(\n)$ and offers the price $p^{\rt(\n)}$; in the case of the rejection, 
it moves to the left node $\lf(\n)$ and offers the price $p^{\lf(\n)}$; this process repeats until reaching the time horizon $T$. 
The pseudo-code of this process is in Alg.~\ref{alg_A}.
The round at which the price of a node $\n\in\T(\A)$ is offered is denoted by $t^{\n}$ (it is equal to the node's depth +1).
Note that each node $\n\in\T(\A)$ uniquely determines the buyer decisions up to the round $t^{\n}-1$. Thus, each buyer strategy $\astr_{1:t}$ is bijectively mapped to a $t$-length path in the tree $\T(\A)$ that starts from the root and goes to a $t$-depth node (and the strategy prices are the ones that are in the nodes lying along this path).

We define, for a pricing tree $\T$, the set of its prices $\PP(\T):=\{p^{\n}\mid \n\in\T \}$  and denote by $\PP(\A) := \PP(\T(\A))$ all prices that can be offered by an algorithm $\A$.
We say that two infinite complete trees $\T_1$ and $\T_2$  are \emph{price equivalent} (and write $\T_1 \cong \T_2$) if the trees have the same node labeling when we naturally match the nodes between the trees (starting from the roots):
i.e., following the same strategy in both trees, the buyer receives the same sequence of prices.

\subsection{Background on pricing algorithms}
\label{subsec_CurrentKnow}

First of all, we remind several classes (sets) of algorithms that were introduced in~\cite{2014-NIPS-Mohri,2017-WWW-Drutsa} and include the definitions of pricing consistency of different type, which are actively used in our work. After that, we briefly overview pricing algorithms from existing studies~\cite{2003-FOCS-Kleinberg,2013-NIPS-Amin,2014-NIPS-Mohri,2017-WWW-Drutsa}.

\underline{\bf Notion of consistency.}
Since the buyer holds a fixed valuation, we could expect that a smart online pricing algorithm should work as follows: after an acceptance (a rejection), it should set only no lower (no higher, resp.) prices than the offered one. Formally, this leads to the definition:
\begin{definition}
	\label{def_Aconsistent}
	An algorithm $\A$ is said to be \emph{consistent}~ \cite{2014-NIPS-Mohri} ($\A$ in the class $\mathbf{C}$) if, for any node $\n\in\T(\A)$,
	$\quad
	p^{\m} \ge p^{\n} \:\:\fa \m\in\RR(\n)
	\quad \hbox{and} \quad
	p^{\m} \le p^{\n} \:\:\fa \m\in\LL(\n).
	$
\end{definition}
The key idea behind a consistent algorithm $\A$ is clear~\cite{2017-WWW-Drutsa}: it explores the valuation domain $[0,1]$ by means of a feasible search interval $[q,q']$ (initialized by $[0,1]$) targeted to locate the valuation $v$. At each round $t$, $\A$ offers a price $p_t\in[q,q']$ and, depending on the buyer's decision, reduces the interval to the right subinterval $[p_t,q']$ (by $q:=p_t$) or the left one $[q,p_t]$ (by $q':=p_t$); at  any moment, $q$ is thus always the last accepted price or $0$, while $q'$ is the last rejected price or $1$. 
The most famous example of a consistent algorithm is the binary search.

\begin{definition}
	\label{def_WC_alg}
	An algorithm $\A$ is said to be \emph{weakly consistent}~\cite{2017-WWW-Drutsa} ($\A$ in the class $\mathbf{WC}$) if, for any node $\n\in\T(\A)$,
	(a) when $\rt(\n)\:\:\hbox{s.t.}\:\: p^{\rt(\n)} \neq p^{\n},$ $\: p^{\m} \ge p^{\n} \:\:\fa \m\in\RR(\n);\:\:$ and,
	(b) when $\lf(\n)\:\:\hbox{s.t.}\:\: p^{\lf(\n)} \neq p^{\n},$ $p^{\m} \le p^{\n} \:\:\fa \m\in\LL(\n).$
	
\end{definition}
Weakly consistent algorithms are similar to consistent ones, but they are additionally able to offer the same price $p$ several times before making a final decision on which of the subintervals $[q,p]$ or $[p,q']$ continue. 
The subclass of WC algorithms that can also wait with the subinterval decision, but the pricing will be the same no matter when a decision is made, is the following.

\begin{definition}
	\label{def_RWC_alg}
	A  weakly consistent algorithm $\A$ is said to be \emph{regular}~\cite{2017-WWW-Drutsa} ($\A$ in the class $\mathbf{RWC}$) if, for any node $\n\in\T(\A)$:
	\begin{itemize}
		\vspace{-0.2cm}
		\item  when $p^{\lf(\n)} = p^{\n} = p^{\rt(\n)}$,
		\label{item_def_RWC_alg_1}
		$
		[p^{\m} = p^{\n} \:\:\fa \m\in\RR(\lf(\n))\cup\LL(\rt(\n))] 
		\:\: \hbox{or} \:\:
		[\LL(\n) \cong \RR(\n)];
		$
		\vspace{-0.2cm}
		\item   when $p^{\lf(\n)} = p^{\n} \neq p^{\rt(\n)}$,
		\label{item_def_RWC_alg_2}
		$
		[p^{\m} = p^{\n} \:\:\fa \m\in\RR(\lf(\n))]  
		\quad\hbox{or}\quad 
		[\RR(\lf(\n)) \cong \RR(\n)] ;
		$
		\vspace{-0.2cm}
		\item  when $p^{\lf(\n)} \neq p^{\n} = p^{\rt(\n)}$,
		\label{item_def_RWC_alg_3}
		$
		[p^{\m} = p^{\n} \:\:\fa \m\in\LL(\rt(\n))]  
		\quad\hbox{or}\quad 
		[\LL(\rt(\n)) \cong \LL(\n)].
		$
	\end{itemize}
\end{definition}

\begin{definition}
	\label{def_C_R_alg}
	An algorithm $\A$ is said to be  \emph{right-consistent}~\cite{2017-WWW-Drutsa} ($\A$ in the class $\mathbf{C_R}$) if, for any  $\n\in\T(\A)$, $p^{\m} \ge p^{\n} \fa \m\in\RR(\n)$.
\end{definition}
Right-consistent algorithms never offer a price lower than the last accepted one, but may offer a price higher than a rejected one (in contrast to consistent algorithms).
These classes are related to each other in the following way: 
$\mathbf{C}
\subset \mathbf{RWC}
\subset \mathbf{WC}$ 
and
$\mathbf{C}
\subset \mathbf{C_R}$.

We will use the following definitions~\cite{2017-WWW-Drutsa} as well.
A buyer strategy $\astr$ is said to be \emph{locally non-losing} w.r.t.\ $v$ and $\A$ if prices higher than $v$ are never accepted\footnote{Note that the optimal strategy of a strategic buyer may not satisfy this property: it is easy to imagine an algorithm that offers the price $1$ at the first round and, if it is accepted, offers the price $0$ all remaining rounds.} (i.e., $a_t = 1$ implies $p_t \le v$).
An algorithm $\A$  is said to be \emph{dense} if the set of its prices $\PP(\A)$ is dense in $[0,1]$ (i.e., $\overline{\PP(\A)} = [0,1]$).

\underline{\bf Background.}
The consistency represents a quite reasonable property, when the buyer is myopic (truthful, i.e., $a_t = 1 \Leftrightarrow p_t \le v$), because a reported buyer decision correctly locates $v$ in $[0,1]$. 
Kleinberg et al.~\cite{2003-FOCS-Kleinberg} showed that the regret of any pricing algorithm against a myopic buyer is lower bounded by $\Omega(\log\log T)$ and proposed a horizon-dependent consistent algorithm, known as \emph{Fast Search} (\emph{FS}), that has tight regret bound in $\Theta(\log\log T)$ against such buyers.

A strategic buyer, incited by surplus maximization, may mislead the seller's consistent algorithm~\cite{2014-NIPS-Amin,2014-NIPS-Mohri}.
To overcome this, Mohri et al.~\cite{2014-NIPS-Mohri} proposed to inject so-called \emph{penalization rounds} (see Def.~\ref{def_PenalNodeSeq}) after each rejection into the algorithm FS and got, in this way, the algorithm PFS with strategic regret bound in $O(\log T\log\log T)$ that outperforms the  algorithm ``Monotone"~\cite{2013-NIPS-Amin} with strategic regret bound in $O(T^{1/2})$. Both algorithms are  horizon-dependent and are not optimal.
\begin{definition}
	\label{def_PenalNodeSeq}
	Nodes $\n_1,\ldots,\n_r\in\T(\A)$ are said to be a ($r$-length) penalization sequence\cite{2014-NIPS-Mohri,2017-WWW-Drutsa} if
	$$
	\n_{i+1} = \lf(\n_i), \qquad p^{\n_{i+1}} = p^{\n_i},  \quad \hbox{and}  \quad \RR(\n_{i+1}) \cong \RR(\n_{i}),  \quad i=1,\ldots,r-1.
	$$
	It is easy to see that a strategic buyer either accepts the price at the first node or rejects this price in all of them, when the discount sequence $\gb$ is decreasing.
\end{definition}

An optimal pricing was found in~\cite{2017-WWW-Drutsa}, where horizon-independent algorithms were studied and the causes of a linear regret in different classes of consistent algorithms were analyzed step-by-step. 
First, the algorithm FES~\cite{2017-WWW-Drutsa} was proposed as a modification of the FS by injecting exploitation rounds after each rejection to obtain a consistent horizon-independent algorithm against truthful buyer with tight regret bound in $\Theta(\log\log T)$.
Second, this pricing was upgraded to the algorithm PRRFES~\cite{2017-WWW-Drutsa} to act against strategic buyers.
Namely, it was shown that there is no no-regret pricing in the class $\mathbf{RWC}$, which comprises, in particular, all consistent horizon-independent algorithms even being modified by penalization rounds. This led to a guess that possibly the left consistency requirement should be relaxed. This guess succeeded in building of the optimal right-consistent algorithm PRRFES with tight strategic regret bound in $\Theta(\log\log T)$, while
the research question on the existence of a no-regret horizon-independent algorithm in the class $\mathbf{WC}$ remained open.

As stated at the beginning of this paper, \emph{our research goals comprise} (a) closing of that open research question; (b) improvement of the best known upper bounds on strategic regret by finding a finer constant factor $C$ of their principal term $C\log\log T$; and (c) generalization of above mentioned results to the cases of strategic buyers whose discounting is not only a geometric progression.

\subsection{Related work}
\label{subsec_RelWork}

Most of studies on online advertising auctions lies in the field of game theory~\cite{2009-Book-Krishna,2007-Book-Nisan}: a large part of them  focused on characterizing different aspects of equilibria, 
and recent ones was devoted (but not limited) to: 
position auctions~\cite{2007-IJIO-Varian,2009-AER-Varian,2014-AER-Varian,2016-WSDM-Charles}, 
different generalizations of second-price auctions~\cite{2009-WWW-Aggarwal,2011-WWW-Celis},
efficiency~\cite{2009-EC-Aggarwal}, 
mechanism expressiveness~\cite{2011-WWW-Dutting}, 
competition across auction platforms~\cite{2013-HBS-Ashlagi}, 
buyer budget~\cite{2014-KDD-Agarwal}, experimental analysis~\cite{2011-EC-Ostrovsky,2013-EC-Thompson,2014-WWW-Noti}, etc.

Studies on revenue maximization were devoted to both the seller revenue solely~\cite{2009-SIGIR-Zhu,2013-IJCAI-He} and different sort of trade-offs either between several auction stakeholders~\cite{2014-WWW-Gomes,2014-WWW-Goel,2014-EC-Bachrach} or between auction characteristics (like revenue monotonicity~\cite{2014-WWW-Goel},  expressivity, and simplicity~\cite{2015-NIPS-Morgenstern}).
The optimization problem was generally reduced to a selection of proper
quality scores for advertisements (for auctions with several advertisers~\cite{2009-SIGIR-Zhu,2013-IJCAI-He}) or
reserve prices for buyers (e.g., in  VCG~\cite{1981-MOR-Myerson}, GSP~\cite{2012-WWW-Lucier}, and others~\cite{2014-WWW-Gomes,2016-WWW-Paes}). The reserve prices, in such setups, usually depend on distributions of buyer bids or valuations and was in turn estimated by machine learning techniques~\cite{2013-IJCAI-He,2014-ECRA-Sun,2016-WWW-Paes}, while alternative approaches learned reserve prices directly~\cite{2014-ICML-Mohri,2015-UAI-Mohri,2016-WWW-Rudolph}.
In contrast to these works, we use an online deterministic learning approach for repeated auctions.

Revenue optimization for repeated
auctions was mainly concentrated on algorithmic reserve prices, that are updated in online fashion over time, and was also known as dynamic pricing.
An extensive survey on this field is presented in~\cite{2015-SORMS-den-Boer}.
Dynamic pricing was studied: 
under game-theoretic view (MFE~\cite{2011-ECOMexch-Iyer,2015-ManagSci-Balseiro}, budget constraints~\cite{2015-ManagSci-Balseiro,2016-EC-Balseiro}, strategic buyer behavior~\cite{2015-EC-Chen}, dynamic mechanisms~\cite{leme2012sequential,2016-EC-Ashlagi}, etc.);
as bandit problems~\cite{2011-COLT-Amin,2015-NIPS-Zoghi,2015-NIPS-Lin} (e.g., UCB-like pricing~\cite{2015-TEC-Babaioff}, bandit feedback models~\cite{2016-JMLR-Weed});
from the buyer side (valuation learning~\cite{2011-ECOMexch-Iyer,2016-JMLR-Weed}, competition between buyers and optimal bidding~\cite{2014-WWW-Hummel,2016-JMLR-Weed}, interaction with several sellers~\cite{2016-ICML-Heidari}, etc.);
from the seller side against several buyers~\cite{2013-SODA-Cesa-Bianchi,2014-KDD-Yuan,2014-SSRN-Kanoria,2016-EC-Roughgarden,2016-NIPS-Feldman};
and a single buyer with stochastic valuation (myopic/truthful~\cite{2003-FOCS-Kleinberg,2011-ICAAMS-Chhabra,2016-SODA-Chawla} and strategic buyers~\cite{2013-NIPS-Amin,2014-NIPS-Amin,2015-NIPS-Mohri,2015-NIPS-Mohri,2016-SSRN-Chen,2017-WWW-Drutsa}, feature-based pricing~\cite{2014-NIPS-Amin,2016-EC-Cohen}, limited supply~\cite{2015-TEC-Babaioff}, etc.).
The most relevant part of these works to ours
are~\cite{2003-FOCS-Kleinberg,2013-NIPS-Amin,2014-NIPS-Mohri,2017-WWW-Drutsa}, where our scenario with a fixed private valuation is considered and whose algorithms are discussed in more details in Sec.~\ref{subsec_CurrentKnow}. 
First, in contrast to~\cite{2003-FOCS-Kleinberg}, we study strategic buyer behavior, whose cumulative surplus may be discounted non-geometrically (unlike in~\cite{2003-FOCS-Kleinberg,2013-NIPS-Amin,2014-NIPS-Mohri,2017-WWW-Drutsa}).
Second, in contrast to~\cite{2013-NIPS-Amin,2014-NIPS-Mohri}, we propose and analyze algorithms that have tight strategic regret bound in $\Theta(\log\log T)$, and, unlike in~\cite{2017-WWW-Drutsa}, one of these algorithms is weakly consistent and never decreases offered prices. Finally, we reduce the factor of the principal term in the strategic regret upper bound from~\cite{2017-WWW-Drutsa}  for the algorithm PRRFES.

\section{Optimizing right-consistent optimal pricing}
\label{sec_RC_OptAlg}
In this section, first, we show that the algorithm PRRFES~\cite{2017-WWW-Drutsa} is able to retain its tight strategic regret bound in $\Theta(\log\log T)$ even against strategic buyers whose surplus is not necessarily discounted geometrically. Second, we provide a finer upper bound for the  PRRFES's strategic regret, that allows to optimize the constant factor $C$ of the principal term $C\log\log T$ of this upper bound by adjusting the number of penalization rounds used in the pricing algorithm. This result allows to obtain a more favorable regret upper bound than in~\cite{2017-WWW-Drutsa}.

For the convenience of readers, we give a short description of the algorithm PRRFES in Appendix~\ref{sec_PRRFES_descr} and its pseudo-code in Alg.~\ref{alg_PRRFES}. 
We begin our regret analysis for discount sequences of general form by proving an analogue of~\cite[Prop.2]{2017-WWW-Drutsa}, which was for a geometric discounting.
Let $\delta_{\n}^{l}:= p^{\n} - \inf_{\m\in\LL(\n)}p^{\m}$ be the left increment~\cite{2014-NIPS-Mohri,2017-WWW-Drutsa}, then 
the following proposition holds.
\begin{proposition}
	\label{prop_reject_bound}
	Let $\gb=\{\g_t\}_{t=1}^\infty$ be a decreasing discount sequence (whose sum  converges), $\A$ be a pricing algorithm, $\n\in\T(\A)$ be a starting node in a $r$-length penalization sequence (see Def.~\ref{def_PenalNodeSeq}), and $r\in\mathbb{N}$ s.t.\ $\g_{t^\n} > \sum_{t=t^\n+r}^\infty\g_t$. If the price $p^{\n}$ offered by the algorithm $\A$ at the node $\n$ is rejected by the strategic buyer, then the following inequality on his valuation $v$ holds:
	\begin{equation}
		\label{prop_reject_bound_eq_1}
		v - p^{\n} < \zeta_{r,\gb,t^\n} \delta_{\n}^{l}, \quad \hbox{where} \quad \zeta_{r,\gb,t} := \frac{\sum_{s=t+r}^\infty\g_s}{\g_{t} - \sum_{s=t+r}^\infty\g_s}. 
	\end{equation}
\end{proposition}
The proof is presented in Appendix~\ref{subsubsec_proof_prop_reject_bound} and is based on ideas similar to the ones in~\cite[Prop.2]{2017-WWW-Drutsa}.
Note that \cite[Prop.2]{2017-WWW-Drutsa} is a particular case of Proposition~\ref{prop_reject_bound}, when $\gb=\{\g^{t-1}\}_{t=1}^{\infty}$ is a geometric discounting for some $\g\in(0,1)$ (then $\zeta_{r,\gb,t}$ becomes $\zeta_{r,\g}$ from \cite[Prop.2]{2017-WWW-Drutsa}).
For this case of geometric discounting, the condition $\g_{t^\n} > \sum_{t=t^\n+r}^\infty\g_t$ on $r$ from Prop.~\ref{prop_reject_bound} becomes $r>\log_\g(1-\g)$. The important property of the latter condition consists in its independence on the time (i.e., round, depth) $t^\n$ of the starting penalization node $\n$. 
This independence property, namely, the property $\exists r\in\mathbb{N}$ s.t.\ $\fa t\in\mathbb{N}: \g_{t} > \sum_{s=t+r}^\infty\g_s$, does not hold for an arbitrary discount sequence $\gb$.  But, in the following lemma, we show that if a discount sequence is geometrically concave, then the above mentioned independence property holds (see the proof in Appendix~\ref{subsubsec_proof_lemma_geom_nonconv_discount}).

\begin{lemma}
	\label{lemma_geom_nonconv_discount}
	Let a decreasing sequence $\gb=\{\g_t\}_{t=1}^\infty$ be geometrically concave, i.e.,  $\g_{t+1}/\g_{t}\ge\g_{t+2}/\g_{t+1} \fa t\in\mathbb{N}$, then (a) there exists $r\in\mathbb{N}$ s.t.\ $\fa t\in\mathbb{N}: \g_{t} > \sum_{s=t+r}^\infty\g_s$; (b) moreover, for any $\varkappa > 0$, there exists $r_\varkappa\in\mathbb{N}$ s.t.\ $\fa t\in\mathbb{N}: \zeta_{r_\varkappa,\gb,t} < \varkappa$.
\end{lemma}
For  geometrically convex discount sequences, the properties in both claims of this lemma may not hold. For instance, consider the telescoping discount $\g_t=1/t(t+1)$ (see Appendix~\ref{subsubsec_proof_telescoping_discount}).

For a right-consistent algorithm $\A$ (and, thus, for the PRRFES as well~\cite{2017-WWW-Drutsa}), the increment $\delta_{\n}^{l}$ in Prop.~\ref{prop_reject_bound} is bounded by the difference between the current node's price $p^{\n}$ and the last accepted price $q$ before reaching this node. 
Hence, the inequality in Eq.~(\ref{prop_reject_bound_eq_1}) provides a guarantee on no-lies at a particular round for certain valuations $v$: the closer an offered price is to the last accepted price the smaller the interval of possible valuations $v$, holding which the strategic buyer may lie on this offer, i.e, the buyer may lie at the $t^\n$-th round only if his valuation $v$ is located in $\big[q, p^{\n} + \zeta_{r,\gb,t^\n} (p^{\n} - q)\big)$.
Using this insight, we can obtain the following theorem, whose proof is presented in Appendix~\ref{subsubsec_proof_th_PRRFES_regret_upper_bound}.

\begin{theorem}
	\label{th_PRRFES_regret_upper_bound}
	Let $\gb=\{\g_t\}_{t=1}^\infty$ be a decreasing discount sequence (whose sum  converges) for which there exist $\varkappa > 0$ and $r_\varkappa\in\mathbb{N}$ s.t.\ $\fa t\in\mathbb{N}: \zeta_{r_\varkappa,\gb,t} < \varkappa$ (the definition of $\zeta_{\cdot,\gb,t}$ is from Eq.~(\ref{prop_reject_bound_eq_1})). If $\A$ is the pricing algorithm PRRFES with $r \ge r_{\varkappa}$ and  the exploitation rate $g(l) = 2^{2^{l}}, l\in\mathbb{Z}_+$, then, for any valuation $v\in [0,1]$ and $T\ge 2$, the strategic regret is upper bounded: 
	\begin{equation}
		\label{th_PRRFES_regret_upper_bound_eq1} 
		\SReg(T,\A, v, \gb)  \le C_{r,\varkappa} (\log_2\log_2 T + 2), \quad \hbox{where} \quad C_{r,\varkappa}:=rv+\frac{(2+\varkappa)^2-1}{2}.
	\end{equation}
\end{theorem}

First, combining Theorem~\ref{th_PRRFES_regret_upper_bound} and Lemma~\ref{lemma_geom_nonconv_discount}, one concludes that \emph{the pricing PRRFES can be effectively applied (with tight regret bound in $\Theta(\log\log T)$), in particular, against strategic buyers with a geometrically concave discount sequence}. Second, the result~\cite[Th.5]{2017-WWW-Drutsa} represents a corollary of Th.~\ref{th_PRRFES_regret_upper_bound}, when we consider 
a geometric discounting $\gb=\{\g^{t-1}\}_{t=1}^{\infty}, \g\in(0,1),$ with parameters $\varkappa=1$ and $r_\varkappa=\lceil\log_{\g}((1-\g)/2)\rceil$.
But, more importantly, Th.~\ref{th_PRRFES_regret_upper_bound} provides a novel regret upper bound (even for a geometric discounting) that can be adjusted, e.g., to reduce the number of penalization rounds or to optimize the constant factor $C_{r,\varkappa}$. 
Let $\gb=\{\g^{t-1}\}_{t=1}^{\infty}, \g\in(0,1)$, then one can easily derive the following dependence between $\varkappa$ and $r_\varkappa$ in order to satisfy the conditions of Th.~\ref{th_PRRFES_regret_upper_bound}: $r_\varkappa\ge\lceil\log_{\g}\big(\frac{\varkappa}{1+\varkappa}(1-\g)\big)\rceil$.
Hence, Th.~\ref{th_PRRFES_regret_upper_bound} allows us to retain the upper bound Eq.~(\ref{th_PRRFES_regret_upper_bound_eq1}) valid and reduce the number of penalization rounds $r$ up to $\lceil\log_{\g}(1-\g)\rceil$. Note that this value is in fact the lower bound for a number of penalization rounds $r$ that satisfies Prop.~\ref{prop_reject_bound} and~\cite[Prop.2]{2017-WWW-Drutsa}. Depending on $\g$, the number of penalization rounds $r$ may be reduced by up to $100/(\log_{1/2}(1-\g)+1)$ per cents; e.g., in integers, we can reduce $r$ from $8$ to $5$ for $\g=0.75$ and  from $72$ to $59$ for $\g=0.95$.

In order to analyze the capacity of possible improvement in optimal factor $C_{r,\varkappa}$, let us bound $r$ by $\log_{\g}\big(\frac{\varkappa}{1+\varkappa}(1-\g)\big) + 1$ and $v$ by $1$, then this upper bound on $C_{r,\varkappa}$ has the following first-order condition w.r.t.\ $\varkappa$: $\varkappa(\varkappa+1)(\varkappa+2) =\ln^{-1}1/\g$, which has only one solution $\varkappa_0$ in $(0,+\infty)$. This solution $\varkappa_0$ monotonically depends on the discount rate $\g$: the closer $\g$ is to $0$ the closer $\varkappa_0$ to $0$, and, vice-versa, $\varkappa_0\to+\infty$ as $\g\to 1$. 
For instance, let us consider the improvement of the bound on the factor $C_{r,\varkappa}$ calculated for the optimal $\varkappa_0$ w.r.t.\ the one for $\varkappa=1$: the factor  is reduced 
by \underline{$33.2\%$}  for $\g=0.05$ with $\varkappa_0\approx 0.137$;
by \underline{$22.9\%$}  for $\g=0.25$ with $\varkappa_0\approx 0.255$;
by \underline{$1.5\%$}  for $\g=0.75$ with $\varkappa_0\approx 0.734$; 
by \underline{$2.8\%$} for $\g=0.95$ with $\varkappa_0\approx 1.815$;
and by \underline{$6.3\%$} for $\g=0.99$ with $\varkappa_0\approx 3.706$.
Thus, we conclude that \emph{Theorem~\ref{th_PRRFES_regret_upper_bound} outperforms the  strategic regret upper bound of~\cite[Th.5]{2017-WWW-Drutsa}}.

\section{Weakly consistent pricing}
\label{sec_WC_alg}
In this section, we, first, generalize the result~\cite[Th.4]{2017-WWW-Drutsa} on the absence of no-regret algorithm in the class $\mathbf{RWC}$ to any discount sequence of the buyer surplus and, moreover, show that any weakly consistent algorithm with double decrease of offered prices has a linear regret. 
This motivates us to hypothesize that there exists a no-regret pricing in $\mathbf{WC}$ which only increases offered prices.
Second, we propose a novel transformation of pricing algorithms and apply it to the algorithm PRRFES obtaining a weakly consistent pricing. 
Finally, we argue that this algorithm is a no-regret one and, moreover, has tight strategic regret bound in $\Theta(\log\log T)$.

\subsection{Weakly consistent algorithms with linear regret}
\label{subsec_WC_alg_lin_reg}
First of all, we isolate the main cause of a linear regret of a wide range of weakly consistent algorithms and formalize it in the following lemma, whose proof is deferred to Appendix~\ref{subsubsec_proof_lemma_WC_SReg_LowerBound}.

\begin{lemma}
	\label{lemma_WC_SReg_LowerBound}
	Let $\gb=\{\g_t\}_{t=1}^\infty$ be a discount sequence and $\A\in\mathbf{WC}$ be a horizon-independent weakly consistent pricing algorithm  s.t.\ the first offered price $p^{\treeroot(\T(\A))}\in(0,1)$. If there exists a path $\tilde{\astr}$ in the tree $\T(\A)$ with the corresponding price sequence $\{\tilde{p}_t\}_{t=1}^{\infty}$ s.t.\ 
	\begin{equation}
	\label{lemma_WC_SReg_LowerBound_eq_1}
	\exists \tilde{t}_0, \tilde{t}_1 \in\mathbb{N}: \quad \tilde{t}_0 \le \tilde{t}_1
	\quad \hbox{and} \quad \tilde{p}_{\tilde{t}_1+1}<\tilde{p}_{\tilde{t}_0}<p^{\treeroot(\T(\A))},
	\end{equation}
	then there exists a valuation $v\in [0,1]$ s.t.\ $\SReg(T,\A, v, \gb) = \Omega(T)$.
\end{lemma}
Note that this lemma holds for any discount sequence and has the following corollary, which is the generalization of~\cite[Th.4]{2017-WWW-Drutsa} to any discounting and whose proof is presented in Appendix~\ref{subsubsec_proof_cor_RWC_SReg_LowerBound}.

\begin{corollary}
	\label{cor_RWC_SReg_LowerBound}
	For any horizon-independent regular weakly consistent pricing algorithm $\A$ and any discount sequence $\gb=\{\g_t\}_{t=1}^\infty$, there exists a valuation $v\in [0,1]$ s.t. $\SReg(T,\A, v, \gb) = \Omega(T)$.
\end{corollary}

The key intuition behind Lemma~\ref{lemma_WC_SReg_LowerBound} consists in the following: the strategic buyer can lie few times to decrease offered prices and, due to (even weak) consistency, receive prices at least on $\ve>0$ lower than his valuation $v$ all the remaining rounds. Note that the buyer is able to mislead a wide range of weakly consistent algorithms: this set of algorithms (that satisfy conditions of Lemma~\ref{lemma_WC_SReg_LowerBound}) is significantly larger than the set $\mathbf{RWC}$ of regular weakly consistent ones. 
But, what if the buyer cannot apply this intuition? After all, there are weakly consistent algorithms that never decrease offered prices.
\emph{We hypothesize thus that there may exist such an algorithm with a sublinear regret}. In Sec.~\ref{subsec_WC_OptAlg}, this hypothesis is  confirmed.

\subsection{Transformation $\pre$}
\label{subsec_PreTransf}

Let us consider a special transformation referred to as $\pre$ and which transforms any pricing algorithm to another one. First, we define this transformation for labeled binary trees. 
\begin{definition}
\label{def_pre_transform}
Given a non-negative real number $q\in\mathbb{R}_+$ and a labeled binary tree $\T_1$, the transformation $\pre:(q,\T_1) \mapsto \T_2$ is such that the labels (i.e., prices) of the tree $\T_2$ are defined recursively  in the following way starting from the root node $\treeroot(\T_2)$ of the tree $\T_2$:
\begin{equation}
\label{eq_def_pre_transform}
p^{\treeroot(\T_2)} := q, \qquad \LL\big(\treeroot(\T_2)\big) \cong \pre\Big(q,\LL\big(\treeroot(\T_1)\big)\Big) \quad \hbox{and} \quad \RR\big(\treeroot(\T_2)\big) \cong \pre\Big(p^{\treeroot(\T_1)},\RR\big(\treeroot(\T_1)\big)\Big).
\end{equation}
\end{definition}
Second, since each pricing algorithm is associated with a complete binary tree $\T(\A)$, the transformation $\pre$ is thus correctly defined for pricing algorithms: namely, $\pre: \mathbb{R}_+\times\ASet \to \ASet$ and $\pre(q,\A_1)$ is the pricing algorithm associated with the tree $\pre(q,\T(\A_1))$.  In Algorithm~\ref{alg_preA}, for better understanding, we provide a reader with a pseudo-code that applies the pricing $\pre(q,\A)$ with given $q\in\mathbb{R}$ and a source pricing $\A\in\ASet$\footnote{We put side-by-side Alg.~\ref{alg_preA} with Alg.~\ref{alg_A} in order to show the difference between the work flow of the transformed pricing $\pre(q,\A)$ and the one of the source pricing $\A$.}.
Informally speaking, this transformation tracks over the nodes in the source algorithm's tree $\T(\A)$, but, being in a current node $\n\in\T(\A)$, it offers the price from one of preceding nodes, where the buyer purchased a good last time (or $q$ if never purchased), instead of offering the price $p^\n$ from the current node $\n$. From the buyer's point of view, the choice between the pricing of the subtrees $\LL(\n)$ and $\RR(\n)$ of a node $\n\in\T(\A)$ should be made at the round \emph{pre}vious to the one where the price $p^\n$ will be offered.
Overall, these intuitions could be used to obtain the following lemma (see the proof in Appendix~\ref{subsubsec_proof_lemma_pre_RC_alg}).

\begin{lemma}
	\label{lemma_pre_RC_alg}
	Let $\A\in\mathbf{C_R}$ be a right-consistent pricing algorithm and $q = \inf\PP(\A)$ be the infimum of the algorithm prices, then the transformed  pricing algorithm $\pre(q,\A)$ is both right-consistent and weakly consistent, i.e., $\pre(q,\A)\in\mathbf{WC}\cap\mathbf{C_R}$.
\end{lemma}
Note that the transformed algorithm $\pre(q,\A)$ for $\A\in\mathbf{C_R}$  is only able to increase prices starting from $q$ and it never decreases them regardless of any buyer  strategy (see Appendix~\ref{subsubsec_proof_lemma_pre_RC_alg}).

\subsection{Weakly consistent optimal pricing}
\label{subsec_WC_OptAlg}
Let us apply the transformation $\pre$ to the pricing algorithm PRRFES and refer to the transformed one as \emph{prePRRFES}.
Formally, the algorithm prePRRFES works in phases initialized by 
the phase index $l:=0$,
 the first offered price at the current phase $q_0:=0$, and 
the iteration parameter $\e_0:=1/2$;
at each phase $l\in\mathbb{Z}_{+}$, it sequentially offers prices $p_{l,k}:= q_{l} + k\e_{l}, k\in\mathbb{Z}_+$ (\underline{exploration}, in contrast to PRRFES, it starts from $k=0$), where
\begin{equation}
\label{eq_prePRRFES_def_round_parameters}
\e_l := \e^2_{l-1} = 2^{-2^l}, \qquad 
\: N_l := \e_{l-1}/\e_{l} = \e^{-1}_{l-1} = 2^{2^{l-1}}, 
\: l\in\mathbb{N};
\end{equation}
if a price $p_{l,k}$ with $k = K_{l}\ge 0$ is rejected,  
(1) it offers this price $p_{l,K_{l}}$ for $r-1$ \underline{penalization} rounds (if one of them is accepted, prePRRFES continues offering $p_{l,k}, k=K_{l}+1,..$ following the Definition~\ref{def_PenalNodeSeq}), 
(2) it offers the price $p_{l,K_{l}}$ for $g(l)$ \underline{exploitation} rounds (buyer decisions made at them do not affect further pricing), and 
(3) prePRRFES goes to the next phase by setting $q_{l+1}:=p_{l,K_{l}}$ and $l:=l+1$.
The pseudo-code of prePRRFES is presented in Alg.~\ref{alg_prePRRFES}, where the lines that differ from the ones of PRRFES (see Alg.~\ref{alg_PRRFES}) are highlighted in {\color{blue} blue}. 
Since PRRFES is a right-consistent algorithm, Lemma~\ref{lemma_pre_RC_alg} implies that prePRRFES is both right-consistent and weakly consistent one.

In this subsection, we will show that prePRRFES being properly configured is, in fact, a no-regret pricing and, moreover, is optimal with tight strategic regret bound in $\Theta(\log\log T)$.
To show this, we follow the methodology of establishing the optimality of the algorithm PRRFES; however, this is not straightforward and requires additional statements (see Prop.~\ref{prop_exploitation_guarantee}) not needed for PRRFES.

In order to simplify further analysis, we assume that the discounting is geometric $\gb=\{\g^{t-1}\}_{t=1}^{\infty}$ from here on in this subsection\footnote{An analysis for non-geometric discount sequences could be done in a similar way as for Theorem~\ref{th_PRRFES_regret_upper_bound}.}.
First, let us consider an analogue of Proposition~\ref{prop_reject_bound} that will be useful to upper bound the strategic regret of the algorithm prePRRFES.

\begin{proposition}
	\label{prop_reject_bound_for_pre}
	Let $\gb=\{\g^{t-1}\}_{t=1}^\infty$ be a discount sequence with $\g\in(0,1)$, $\A$ be a pricing algorithm, $\n\in\T(\A)$ be a starting node in a $r$-length penalization sequence (see Def.~\ref{def_PenalNodeSeq}), all prices after $r$ rejections are no lower than $p^\n$ (i.e., $p^{\n}\le p^{\m}  \: \fa\m\in\LL(\lf^{r-1}(\n))$), and $r > \log_\g(1-\g^2)$. If the price $p^{\n}$ offered by the algorithm $\A$ at the node $\n$ is rejected by the strategic buyer, then the following inequality on his valuation $v$ holds:
	\begin{equation}
	\label{prop_reject_bound_for_pre_eq_1}
	v - p^{\rt(\n)} < \eta_{r,\g} (p^{\rt(\n)} - p^{\n}), \quad \hbox{where} \quad \eta_{r,\g} := \frac{\g^r+\g - 1}{1-\g^2-\g^r}. 
	\end{equation}
\end{proposition}
The proof is given in Appendix~\ref{subsubsec_proof_prop_reject_bound_for_pre}.
Note that, similarly to Eq.~(\ref{prop_reject_bound_eq_1}), the inequality in Eq.~(\ref{prop_reject_bound_for_pre_eq_1}) bounds the deviation of the buyer's valuation $v$ from the price at some node $\rt(\n)$ by some increment $p^{\rt(\n)} - p^{\n}$. But, in contrast to Eq.~(\ref{prop_reject_bound_eq_1}),
this bounding occurs when the buyer rejects the price $p^{\n}$ offered previously to the one $p^{\rt(\n)}$ which is used as the reference price of the valuation's deviation.

As we show in the proof of Theorem~\ref{th_prePRRFES_regret_upper_bound}, Prop.~\ref{prop_reject_bound_for_pre} allows us to obtain an upper bound for the number of exploring steps at each phase of the algorithm prePRRFES (like Prop.~\ref{prop_reject_bound} is used for the algorithm PRRFES). However, this is \emph{not enough to directly apply} the methodology of the proofs of \cite[Th.3, Th.5]{2017-WWW-Drutsa} and Theorem~\ref{th_PRRFES_regret_upper_bound} to bound the strategic regret, because, in contrast to the PRRFES, during exploitation rounds, the algorithm prePRRFES offers the price $p_{l, K_{l}}$ that has not been earlier accepted by the strategic buyer (hence, there is no evidence to guarantee his acceptance during the exploitation). Namely, since the buyer's decision $a_t$ made at an exploitation round $t$ does not affect the algorithm's pricing in the subsequent rounds $t'>t$, the strategic buyer acts truthfully at this round $t$, i.e., $a_t=\Ind_{\{p_t\le v\}}$. For the PRRFES, we knew that the price $p_t$ was accepted in a previous round $t'<t$ (or $p_t=0$), but, for the prePRRFES, one has to specially guarantee the acceptance of the price $p_t$ at the exploitation round $t$ in the following proposition.

\begin{proposition}
	\label{prop_exploitation_guarantee}
	Let $\gb=\{\g^{t-1}\}_{t=1}^\infty$ be a discount sequence with $\g\in(0,1)$, $\A$ be a pricing algorithm, and 
	$\n\in\T(\A)$ be a starting node in a $r$-length penalization sequence (see Def.~\ref{def_PenalNodeSeq}), which is followed by $G$ exploitation rounds offering the price $p^\n$ starting from the node $\lf^r(\n)$. If 
	$r < \log_\g(1-\g)$, $G>\log_\g\big(1 - (1-\g)\g^{-r}\big)$, 
	$T\ge t^\n+r+G-1$, and the buyer valuation $v$ is higher than $p^\n$ and lower than any price in the right subtree $\RR(\n)$ of the node $\n$, i.e., $v < p \fa p\in\PP(\RR(\n))$, 
	then the strategic buyer rejects the price $p^\n$ at the round $t^\n$.
\end{proposition}
The proof is presented in Appendix~\ref{subsubsec_proof_prop_exploitation_guarantee}. Additionally to the claim of Proposition~\ref{prop_exploitation_guarantee}, note that, from the definitions of penalization and exploitation rounds, it follows that, if the strategic buyer rejects the price $p^\n$ at the round $t^\n$,  he rejects this price $p^\n$ at the rounds $t^\n+1,\ldots,t^\n+r-1$ as well and accepts it at the rounds $t^\n+r,\ldots,t^\n+r+G-1$.
Note that, since $r\ge 1$ (otherwise, there is no node $\n$ and the right subtree $\RR(\n)$), the condition $r < \log_\g(1-\g)$ makes  Prop.~\ref{prop_exploitation_guarantee} meaningful only in the case of $\g>1/2$. This is consistent with a clear intuition that, having $\g\le1/2$, the discount $\g^{t-1}$ at a round $t$ is no lower than the sum of all discounts in all possible subsequent rounds $\g^t/(1-\g)$, and the strategic buyer prefers thus to purchase a good for a price $p_t$ at the $t$-th round, rather than many goods for a no lower price in all subsequent rounds.

In order to use both Prop.~\ref{prop_reject_bound_for_pre} and Prop.~\ref{prop_exploitation_guarantee},  the number of penalization rounds $r$ is required to be in $\big(\log_\g(1-\g^2), \log_\g(1-\g)\big)$. We restrict $\g$ by the condition $(1+\g)\g>1$ in order to  have the length of this interval larger than $1$ and guarantee thus existence of a natural number in it (since $r\in\mathbb{N}$). This restriction implies that $\g$ should be larger than $(\sqrt{5}-1)/2$. For such discount rates, the following lemma (with the proof in Appendix~\ref{subsubsec_proof_lemma_penalty_number_estimate}) provides values for $r$ and $G$ s.t.\ Prop.~\ref{prop_reject_bound_for_pre} and Prop.~\ref{prop_exploitation_guarantee} hold and $\eta_{r,\g}$ (from Eq.~(\ref{prop_reject_bound_for_pre_eq_1})) is bounded by some positive number $\varkappa > 0$.

\begin{lemma}
	\label{lemma_penalty_number_estimate}
	Let the discount rate $\g\in\big((\sqrt{5}-1)/2,1\big)$, a constant $\varkappa>(1-\g)/(\g^2+\g-1)$, the number of penalization rounds $r = \lceil r_{\g,\varkappa} \rceil$, and the number of exploitation rounds $G\ge G_{\g,\varkappa}$, where 
	\begin{equation}
	\label{lemma_penalty_number_estimate_eq_1}
	r_{\g,\varkappa}:=\log_\g\left((1-\g)\Big(1+\frac{\varkappa}{1+\varkappa}\g\Big)\right) 
	\quad \hbox{and} \quad
	G_{\g,\varkappa}:= \log_\g\left(1 - \Big(1+\frac{\varkappa}{1+\varkappa}\g\Big)^{-1}\g^{-1}\right).
	\end{equation}
	Then the conditions of Prop.~\ref{prop_reject_bound_for_pre} and Prop.~\ref{prop_exploitation_guarantee} hold, and  $\eta_{r,\g} \le \varkappa$.
\end{lemma}

Now we ready to obtain an upper bound for the prePRRFES by proving the following theorem.
\begin{theorem}
	\label{th_prePRRFES_regret_upper_bound}
	Let $\gb=\{\g^{t-1}\}_{t=1}^\infty$ be a discount sequence with $\g\in\big((\sqrt{5}-1)/2,1\big)$,
	a constant $\varkappa>(1-\g)/(\g^2+\g-1)$, while the constants $r_{\g,\varkappa}$ and $G_{\g,\varkappa}$ be from Eq.~(\ref{lemma_penalty_number_estimate_eq_1}).
	If $\A$ is the pricing algorithm prePRRFES with  $r = \lceil r_{\g,\varkappa} \rceil$ and  the exploitation rate  $g(l) = \max\{2^{2^{l}},\lceil G_{\g,\varkappa}\rceil\}, l\in\mathbb{Z}_+$, then, for any valuation $v\in [0,1]$ and $T\ge 2$, the strategic regret is upper bounded: 
	\begin{equation}
	\label{th_prePRRFES_regret_upper_bound_eq1} 
	\SReg(T,\A, v, \gb)  \le \left(rv+\frac{(1+\varkappa)}{2}(2+\max\{2,\lceil G_{\g,\varkappa}\rceil\}+\varkappa)\right)(\log_2\log_2 T + 2)+ \frac{\lceil G_{\g,\varkappa}\rceil}{2} - 1.
	\end{equation}
\end{theorem}
The proof of this theorem  is presented in Appendix~\ref{subsubsec_proof_th_prePRRFES_regret_upper_bound} and is based on the methodology of the proof of Th.~\ref{th_PRRFES_regret_upper_bound}, but requires special modifications as discussed above. 
Theorem~\ref{th_prePRRFES_regret_upper_bound} confirms our hypothesis on the existence of a no-regret algorithm in the class $\mathbf{WC}$ and \emph{closes thus the corresponding open research question~\cite{2017-WWW-Drutsa}}.
An attentive reader may note that the pricing prePRRFES has the following drawback: this algorithm being applied against a myopic (truthful) buyer will incur a linear regret (in contrast to the source PRRFES). 
But we feel that this is the price we have to pay in order to construct a horizon-independent optimal algorithm that offers prices in a consistent manner (i.e., never revise prices that was previously reduced as it did by the PRRFES).

\section{Conclusions}
\label{sec_Conclusions}
We studied horizon-independent online learning (discrete) algorithms in the scenario of repeated posted-price auctions with a strategic buyer that holds a fixed private valuation.
First, we closed the open research question on the existence of a no-regret horizon-independent weakly consistent algorithm by proposing a novel algorithm that never decreases offered prices and can be applied against strategic buyers with a tight regret bound in $\Theta(\log\log T)$.
Second, we provided an upper bound on strategic regret of the algorithm PRRFES, that allows to optimize the constant factor $C$ of its principal term $C\log\log T$, outperforming thus the previously best known upper bounds.
Finally, we generalized the previously known  lower and upper bounds on strategic regret to classes of discount sequences that are wider than geometric progressions.

\appendix
\numberwithin{equation}{section}
\numberwithin{algorithm}{section}
\numberwithin{figure}{section}
\numberwithin{table}{section}

\section{Missed proofs}

\subsection{Missed proofs from Section~\ref{sec_RC_OptAlg}}

\subsubsection{Proof of Proposition~\ref{prop_reject_bound}}
\label{subsubsec_proof_prop_reject_bound}
\begin{proof}
	For each node $\m\in\T(\A)$, let $S(\m)$ be the surplus obtained by the buyer when playing an
	optimal strategy against $\A$ after reaching the node $\m$. Since  the price $p^{\n}$ is rejected at the node $\n$ by the strategic buyer, there following inequality on surpluses holds:
	\begin{equation}
	\label{prop_reject_bound_proof_eq_0}
	\g_{t^\n}(v-p^{\n}) + S(\rt(\n)) <  S(\lf(\n)).
	\end{equation}
	First, for the case $r>1$, we show that rejection of the price $p^{\n}$ at the node $\n$ implies rejection of this price at the subsequent penalization nodes $\lf^s(\n), s=1,\ldots,r-1$, as well. 
	Indeed, let us assume the contrary: the strategic buyer accepts the price at the node $\lf^s(\n)$ for some $s=1,\ldots,r-1$ (let $s$ be the smallest one). Hence, the optimal strategy passes through the right subtree $\RR(\lf^s(\n))$, which is price equivalent to the tree $\RR(\n)$, i.e., $\RR(\lf^s(\n)) \cong \RR(\n)$ (by Def.~\ref{def_PenalNodeSeq} of a penalization sequence). Let $\{\tilde a_t\}_{t=t^n+s+1}^{\infty}$ be the part of this optimal strategy within the subtree $\RR(\lf^s(\n))$ (with the corresponding sequence of prices $\{\tilde p_t\}_{t=t^n+s+1}^{\infty}$), then we
	consider the strategy $\{\hat a_t\}_{t=t^n+1}^{\infty}$ with the same sequence of decisions, but made after acceptance of the price $p^{\n}$ at the node $\n$: $\hat a_t := \tilde a_{t+s}$. Since $\RR(\lf^s(\n)) \cong \RR(\n)$, the corresponding prices will be the same: $\hat p_t = \tilde p_{t+s}$. Hence, we have 
	\begin{equation}
	\begin{split}
	\label{prop_reject_bound_proof_eq_0a}
	S(\lf(\n)) = S(\lf^s(\n)) = \g_{t^\n+s}\left(v-p^{\lf^s(\n)}\right) + S(\rt(\lf^s(\n))) 
	= \g_{t^\n+s}\left(v-p^{\lf^s(\n)}\right) + \!\!\!\!\!\sum_{t=t^\n+s+1}^{\infty}\!\!\!\!\!\g_{t}\tilde a_t(v-\tilde p_t) = 
	\\ \g_{t^\n+s}(v-p^{\n}) + \!\!\!\sum_{t=t^\n+1}^{\infty}\!\!\!\g_{t+s}\hat a_t(v-\hat p_t) < \g_{t^\n}(v-p^{\n}) + \!\!\!\sum_{t=t^\n+1}^{\infty}\!\!\!\g_{t}\hat a_t(v-\hat p_t)
	\le \g_{t^\n}(v-p^{\n}) + S(\rt(\n)),
	\end{split}
	\end{equation}
	where we used, in the first inequality, the fact that the discount sequence $\gb$ is decreasing and,   in the second one, that $\{\hat a_t\}_{t=t^n+1}^{\infty}$ generates surplus in $\RR(\n)$ at most $S(\rt(\n))$. In Eq.~(\ref{prop_reject_bound_proof_eq_0a}), we obtain a contradiction to Eq.~(\ref{prop_reject_bound_proof_eq_0}).
	Therefore, the following inequality holds:
	\begin{equation}
	\label{prop_reject_bound_proof_eq_1}
	\g_{t^\n}(v-p^{\n}) + S(\rt(\n)) <  S(\lf(\n)) = S(\lf^{r}(\n)).
	\end{equation}
	The surplus $S(\rt(\n))$ is lower bounded by $0$, while the left subtree's surplus $S(\lf^r(\n))$ can be upper bounded as follows (using $p^{\n}-p^{\m}\le \delta_{\n}^{l} \: \fa\m\in\LL(\n)$):
	$$
	S(\lf^r(\n)) \le \sum\limits_{t=t^\n+r}^T\g_{t}(v - p^{\n} + \delta_{\n}^{l}) < \sum\limits_{t=t^\n+r}^\infty\g_{t}(v - p^{\n} + \delta_{\n}^{l}),
	$$
	We plug these bounds in Eq.~(\ref{prop_reject_bound_proof_eq_1}) and obtain
	$$
	(v - p^{\n})\left(\g_{t^{\n}} - \sum\limits_{t=t^\n+r}^\infty\g_{t}\right)    <  \sum\limits_{t=t^\n+r}^\infty\g_{t} \delta_{\n}^{l},
	$$
	that implies Eq.~(\ref{prop_reject_bound_eq_1}), since $r$ s.t.\ $\g_{t^\n} > \sum_{t=t^\n+r}^\infty\g_t$. 
\end{proof}

\subsubsection{Proof of Lemma~\ref{lemma_geom_nonconv_discount}}
\label{subsubsec_proof_lemma_geom_nonconv_discount}
\begin{proof}
	Let $\alpha_{t+1} := \g_{t+1}/\g_{t}, t\in\mathbb{N}$, then, from the non-convexity, we have $\alpha_{t+1}\ge\alpha_{t+2} \fa t\in\mathbb{N}$. In particular, $\alpha_{t+1}\le\alpha_{2}<1, t\in\mathbb{N},$ since $\gb$ is decreasing. So, for any $t,t'\in\mathbb{N}$, 
	$$
	\g_{t+t'} = \alpha_{t+t'}\alpha_{t+t'-1}\ldots\alpha_{t+1}\g_t \le \alpha_t^{t'}\g_t\le\alpha_2^{t'} \g_t.
	$$
	Hence, 
	$$\sum_{s=t+r}^\infty\g_s \le \sum_{s=t+r}^\infty\alpha_2^{s-t}\g_{t} = \frac{\alpha_2^{r}}{1-\alpha_2}\g_t.$$ 
	Taking $r>\log_{\alpha_2}(1-\alpha_2)$ and $r_\varkappa>\log_{\alpha_2}\big(\varkappa(1-\alpha_2)/(1+\varkappa)\big)$, we obtain both claims of the lemma.
\end{proof}

\subsubsection{Telescoping discount sequence $\g_t=1/t(t+1)$ does not satisfy Prop.~\ref{prop_reject_bound}}
\label{subsubsec_proof_telescoping_discount}
Note that this discount sequence is geometrically convex, i.e., $\g_{t+1}/\g_{t}\le\g_{t+2}/\g_{t+1} \fa t\in\mathbb{N}$.

Let us show that for $\g_t=\frac{1}{t(t+1)}$ the property $\exists r\in\mathbb{N}$ s.t.\ $\fa t\in\mathbb{N}: \g_{t} > \sum_{s=t+r}^\infty\g_s$ does not hold. Indeed, assume the contrary: let $r$ be s.t.\ $\fa t\in\mathbb{N}: \g_{t} > \sum_{s=t+r}^\infty\g_s$. We have $\sum_{s=t+r}^\infty\g_s = 1/(t+r-1)$, which implies $\fa t\in\mathbb{N}: \frac{1}{t(t+1)} > \frac{1}{t+r-1}$ or, equivalently, $t+r-1 > t^2+t$, i.e., $r>t^2+1$. Take $t = r$ and get a contradiction in $r>r^2+1$.

\subsubsection{Proof of Theorem~\ref{th_PRRFES_regret_upper_bound}}
\label{subsubsec_proof_th_PRRFES_regret_upper_bound}

\begin{proof}
	The proof is fairly similar to the one of \cite[Th.5]{2017-WWW-Drutsa}.
	So, let $L$ be the number of phases conducted by the algorithm during $T$ rounds, then we decompose the total regret over $T$ rounds into the sum of the phases' regrets: $\SReg(T,\A, v, \gb) = \sum_{l=0}^{L}R_l$. For the regret $R_l$ at each phase except the last one, the following identity holds:
	\begin{equation}
	\label{th_PRRFES_regret_upper_bound_proof_eq1}
	R_l =  \sum\limits_{k=1}^{K_l} (v-p_{l,k}) + rv + g(l)(v-p_{l,K_{l}}), \quad l=0,\ldots,L-1,
	\end{equation} 
	where the first, second, and third terms correspond to the exploration rounds with acceptance, the reject-penalization rounds, and the exploitation rounds\footnote{Note that the prices at the exploitation rounds $p_{l, K_{l}}$ are equal to either $0$ or an earlier accepted price, and are thus accepted by the strategic buyers (since the buyer's decisions at these rounds do not affect further pricing of the algorithm PRRFES).}, respectively.
	First, note that the optimal strategy is locally non-losing one for $\A\in\mathbf{C_R}$ (see the discussion after \cite[Lemma~1]{2017-WWW-Drutsa}). Hence, since the price $p_{l,K_{l}}$ is $0$ or has been accepted, we have  $p_{l,K_{l}} \le v$. 
	Second, since the price $p_{l,K_{l} + 1}$ is rejected, we have $v - p_{l,K_{l} + 1} < \varkappa(p_{l,K_{l} + 1} - p_{l,K_{l}})=\varkappa\e_l$ (by Proposition~\ref{prop_reject_bound} since $\zeta_{r,\gb,t}<\varkappa$ for $r\ge r_{\varkappa}$ and any $t\in\mathbb{N}$). Hence, the valuation 
	$v\in\big[p_{l,K_{l}}, p_{l,K_{l}} + (1+\varkappa)\e_{l}\big)$ and all accepted prices $p_{l+1,k}, \fa k\le K_{l+1}$, from the next phase $l+1$ satisfy: 
	$$
	p_{l+1,k}\in[q_{l+1},v)\subseteq \big[p_{l,K_{l}}, p_{l,K_{l}} + (1+\varkappa)\e_{l}\big) \quad \fa k\le K_{l+1},
	$$ 
	because any accepted price has to be lower than the valuation~$v$ for the strategic buyer (whose optimal strategy is locally non-losing one, as we stated above). This infers $K_{l+1} <(1+\varkappa)N_{l+1}\le\big\lceil(1+\varkappa)N_{l+1}\big\rceil=:N_{l+1,\varkappa}$. Therefore, for the phases $l=1,\ldots,L$, we have:
	\begin{equation*}
	v - p_{l,K_{l}} < (1+\varkappa)\e_{l}; \qquad
	v-p_{l,k} < \e_{l}\big((1+\varkappa)N_{l}-k\big)  \fa k\in\mathbb{Z}_{N_{l,\varkappa}};
	\end{equation*}
	and
	\begin{equation*}
	\begin{split}
	\sum\limits_{k=1}^{K_{l}}(v-p_{l,k}) < \e_{l}\sum\limits_{k=1}^{N_{l,\varkappa}-1}\big((1+\varkappa)N_{l}-k\big) = \e_{l}\frac{N_{l,\varkappa}-1}{2}\big(2(1+\varkappa)N_l-N_{l,\varkappa}\big) \le
	\\
	\le \e_{l}\frac{(1+\varkappa)N_l}{2}(1+\varkappa)N_l = \frac{(1+\varkappa)^2}{2}N_l\cdot N_l\e_{l} = \frac{(1+\varkappa)^2}{2}N_l\cdot \e_{l-1} = \frac{(1+\varkappa)^2}{2},
	\end{split}
	\end{equation*} 
	where we used the definitions of $N_l$ and $\e_l$ (i.e., $N_l\e_l = \e_{l-1}$ and $N_l = \e_l^{-1}$).
	For the zeroth phase $l=0$, one has trivial bound 
	$\sum_{k=1}^{K_{0}}(v-p_{0,k}) \le 1/2$.
	Hence, by definition of the exploitation rate $g(l)$, we have $g(l)=\e_l^{-1}$ and, thus,
	\begin{equation}
	\label{th_PRRFES_regret_upper_bound_proof_eq2} 
	R_l \le  \frac{(1+\varkappa)^2}{2} + rv + g(l) \cdot (1+\varkappa)\e_{l} \le  rv+\frac{(2+\varkappa)^2-1}{2}, \quad l=0,\ldots,L-1.
	\end{equation} 
	
	Moreover, this inequality holds for the $L$-th phase, since it differs from the other ones only in possible absence of some rounds (reject-penalization or exploitation ones). Namely, for the $L$-th phase, we have:
	\begin{equation}
	\label{th_PRRFES_regret_upper_bound_proof_eq3} 
	R_L =  \sum\limits_{k=1}^{K_L} (v-p_{L,k}) + r_Lv + g_L(L)(v-p_{L,K_{L}}),
	\end{equation}
	where $r_L$ is the actual number of reject-penalization rounds and $g_L(L)$ is the actual number of exploitation ones in the last phase. Since $r_L\le r$ and $g_L(L)\le g(L)$,  
	the right-hand side of Eq.~(\ref{th_PRRFES_regret_upper_bound_proof_eq3}) is upper-bounded by the right-hand side of Eq.~(\ref{th_PRRFES_regret_upper_bound_proof_eq1}) with $l=L$, which is in turn upper-bounded by the right-hand side of Eq.~(\ref{th_PRRFES_regret_upper_bound_proof_eq2}).
	Finally, one has
	\begin{equation*}
	\SReg(T,\A, v, \gb) = \sum_{l=0}^{L}R_l \le \left(rv+\frac{(2+\varkappa)^2-1}{2}\right)(L+1).
	\end{equation*} 
	Thus, one needs only to estimate the number of phases $L$ by the number of rounds $T$. So, for $2\le T\le 2 + r + g(0)$, we have $L=0$ or $1$ and thus $L+1\le 2 \le \log_2\log_2 T + 2$. For $T \ge 2 + r + g(0)$, we have $T = \sum_{l=0}^{L-1}(K_l + r + g(l)) + K_L + r_L + g_L(L)\ge g(L-1)$ with $L>0$.
	Hence, $g(L-1) = 2^{2^{L-1}}\le T$, which is equivalent to $L \le \log_2\log_2 T + 1$. Summarizing, we get Eq.~(\ref{th_PRRFES_regret_upper_bound_eq1}).
\end{proof}

\subsection{Missed proofs from Section~\ref{subsec_WC_alg_lin_reg}}

\subsubsection{Proof of Lemma~\ref{lemma_WC_SReg_LowerBound}}
\label{subsubsec_proof_lemma_WC_SReg_LowerBound}
\begin{proof}
	Let us denote the first offered price as $p_1:=p^{\treeroot(\T(\A))}$\footnote{Note that $p_1$ is the first element in a price sequence of any buyer strategy for a given $\A$.} and decompose the set of all buyer strategies (i.e., paths in the tree $\T(\A)$) into three sets $B_0 \sqcup B_{-} \sqcup B_{+}$:
	\begin{itemize}
		\item  $B_0$ contains strategies whose price sequences $\{p_t\}_{t=1}^{\infty}$ are constant: $p_t=p_1 \fa t\in\mathbb{N}$;
		\item  for a strategy from $B_{-}$, the price sequence $\{p_t\}_{t=1}^{\infty}$ has the form: $\exists t_0\in\mathbb{N}$ s.t.\ $p_{t_0+1}<p_{t_0}$ and $p_t=p_1, t=1,\ldots,t_0$;
		\item  for a strategy from $B_{+}$, its price sequence $\{p_t\}_{t=1}^{\infty}$ has the form: $\exists t_0\in\mathbb{N}$ s.t.\ $p_{t_0+1}>p_{t_0}$ and $p_t=p_1, t=1,\ldots,t_0$.
	\end{itemize}
	Note that $B_{-}\neq \varnothing$ since it contains the strategy $\tilde{\astr}$ whose price sequence is defined in Eq.~(\ref{lemma_WC_SReg_LowerBound_eq_1}). Based on this strategy $\tilde{\astr}$, let us consider the strategy $\hat{\astr}$ s.t.\ it coincides with $\tilde{\astr}$ at up to  the $\tilde{t}_1$-th round, i.e., $\hat{\astr}_t := \tilde{\astr}_t \fa t \le\tilde{t}_1$, and $\hat{\astr}_t := 1 \fa t> \tilde{t}_1$. 
	It easy to see that $\hat{\astr}\in B_{-}$.
	We denote the corresponding price sequence by  $\{\hat{p}_t\}_{t=1}^{\infty}$.
	
	Let us denote $\Delta = p_1 - \hat{p}_{\tilde{t}_0}>0$, then, $\fa t \ge \tilde{t}_1$, $\hat{p}_t\le \hat{p}_{\tilde{t}_0} = p_1 - \Delta$ (due to the weak consistency of the algorithm $\A$\footnote{In fact, it easy to derive (using the weak consistency) that $\hat{p}_t\le \hat{p}_{\tilde{t}_0} \fa t \ge \tilde{t}_0$. Because, otherwise, if $\exists t': \tilde{t}_0<t'<\tilde{t}_1$ s.t.\ $\hat{p}_{t'}>\hat{p}_{\tilde{t}_0}$, which implies $\hat{p}_{t}\ge\hat{p}_{\tilde{t}_0} \fa t>t'$ and contradicts to $\hat{p}_{\tilde{t}_1}<\hat{p}_{\tilde{t}_0}$.}). Hence, on the one hand, the surplus of the strategy $\hat{\astr}$ followed by a buyer with the valuation $v_{\ve}:= p_1 +\ve$ can be lower bounded in the following way:
	\begin{equation}
	\label{eq_lemma_WC_SReg_LowerBound_proof_1}
	\Sur_{\gb}(T,\A,v_{\ve},\hat{\astr}) \ge \sum\limits_{t=\tilde{t}_1+1}^{T}\g_{t} (\Delta + \ve) \qquad \fa T>\tilde{t}_1.
	\end{equation}
	
	On the other hand, one can upper bound the surplus of a strategy $\astr\in B_{+}$ followed by a buyer with the valuation $v_{\ve}$ since the price sequence corresponding to $\astr$ satisfies $p_t \ge p_1 \fa t\in\mathbb{N}$:
	\begin{equation}
	\label{eq_lemma_WC_SReg_LowerBound_proof_2}
	\Sur_{\gb}(T,\A,v_{\ve},\astr) \le \sum\limits_{t=1}^{T}\g_{t} \ve \quad \fa \astr\in B_{+} \fa T>0.
	\end{equation}
	Let  
	$$
	\ve_0 :=\min\left\{ \Delta  \frac{\g_{\tilde{t}_1+1}}{\sum_{t=1}^{\tilde{t}_1}\g_{t}}, 1-p_1\right\},
	$$
	then, $\fa \ve\in(0,\ve_0)$, first, $v_{\ve}\in(0,1)$ and, second,
	$$
	\ve < \Delta \frac{\sum_{t=\tilde{t}_1+1}^{T}\g_{t}}{\sum_{t=1}^{\tilde{t}_1}\g_{t}} \qquad \fa T>\tilde{t}_1.
	$$
	Therefore, the right-hand side of Eq.~(\ref{eq_lemma_WC_SReg_LowerBound_proof_1}) is larger than the one of Eq.~(\ref{eq_lemma_WC_SReg_LowerBound_proof_2}), which implies that 
	$$
	\Sur_{\gb}(T,\A,v_{\ve},\astr)<\Sur_{\gb}(T,\A,v_{\ve},\hat{\astr}) \qquad \fa \astr\in B_{+}.
	$$ 
	
	Thus, we showed that, for $T>\tilde{t}_1$, there exists a strategy  in $B_{-}$ (namely, $\hat{\astr}$) that is better (in terms of discounted surplus) than any strategy in $B_{+}$ for the buyer with the valuation $v_{\ve} = p_1+\ve, \ve\in(0,\ve_0)$. Therefore, the optimal strategy $\astr^{\Opt}$ must belong to either $B_0$ or $B_{-}$ for $T>\tilde{t}_1$. But, for any strategy $\astr$ from $B_0\cup B_{-}$, one can lower bound the regret by
	$$
	\Reg(T,\A, v_\ve, \astr) \ge \sum\limits_{t:a_t=0} v_\ve + \sum\limits_{t:a_t=1} (v_\ve - p_1) \ge T\ve,
	$$
	and, hence, the strategic regret: $\SReg(T,\A, v_\ve, \gb)\ge T\ve$ for $T>\tilde{t}_1$.
	This lower bound is $\Omega(T)$ since $\ve$ and $\tilde{t}_1$ are independent of $T$. 
\end{proof}

\subsubsection{Proof of Corollary~\ref{cor_RWC_SReg_LowerBound}}
\label{subsubsec_proof_cor_RWC_SReg_LowerBound}

\begin{proof}
	If the algorithm $\A$ is not dense, then the theorem holds since any non-dense horizon-independent regular weakly consistent algorithm has linear strategic regret (see \cite[Cor.1]{2017-WWW-Drutsa}). First, let us consider the case when the first offered price $p_1:=p^{\treeroot(\T(\A))}\in(0,1)$ and show existence of a path $\tilde{\astr}$ in the tree $\T(\A)$ that satisfies Eq.~(\ref{lemma_WC_SReg_LowerBound_eq_1}) from Lemma~\ref{lemma_WC_SReg_LowerBound}.
	
	Indeed, since $\A$ is dense there exists a node $\n\in\T(\A)$ s.t.\ $p^\n\in(0,p_1)$; let us take the one with the smallest depth $t^\n$, denote $p':=p^\n; t':=t^\n$, and consider the path $\hat{\astr}_{1:t'-1}$ from the root to this node $\n$. For the corresponding price sequence $\{\hat{p}_t\}_{t=1}^{t'}$, the following holds:
	\begin{itemize}
		\item $\hat{p}_t \le p_1 \fa t \le t'$ due to the weak consistency of the algorithm $\A$;
		\item $\hat{p}_t\in\{0,p_1\} \fa t < t'$ due to the choice of the node $\n$ with minimal $t^\n$.
	\end{itemize}
	Since the algorithm $\A$ is regular weakly consistent, for any path $\astr$ from the root s.t.\ its price sequence $\{p_t\}_{t=1}^{\infty}$ contains a price lower than $p_1$, the price sequence $\{p_t\}_{t=1}^{\infty}$  must be similar to $\{\hat{p}_t\}_{t=1}^{t'}$ at the beginning. Namely, there exists a node $\m\in\T(\A)$ s.t.\ the path $\astr$ passes through this node $\m$, $p^{\m} = p_{t^\m} = p'$, and $p_t\in\{0,p_1\} \fa t < t^\m$. Moreover, $\T(\m)\cong\T(\n)$ since $\A\in\mathbf{RWC}$ as well.
	Hence, if $\PP(\T(\n))\cap(0,p')=\varnothing$, then $\PP(\T(\A))\cap(0,p')=\varnothing$, that contradicts to the density of the algorithm $\A$. Therefore, there exists a node $\hat{\n}\in\T(\n)$ s.t.\ $p^{\hat{\n}}\in(0,p')$. Continuing the path $\hat{\astr}_{1:t'-1}$ to this node $\hat{\n}$, one gets the desired path $\tilde{\astr}$ in the tree $\T(\A)$ that satisfies Eq.~(\ref{lemma_WC_SReg_LowerBound_eq_1}) from Lemma~\ref{lemma_WC_SReg_LowerBound}, which implies linear strategic regret for the algorithm $\A$ in the case $p_1\in(0,1)$.
	
	Let us consider the case of $p_1=0$ or $1$. 
	Since $\A$ is dense, then, there exists a node $\n\in\T(\A)$ such that $p^{\n}\in(0,1)$; we denote by $\tilde{\n}$ the one among them with the smallest depth $t^{\n}$. 
	So, the problem of strategic regret estimation reduces to the previously considered case of $0<p_1<1$ and resolves by replacing $p_1$ with $p^{\tilde{\n}}$ in our reasoning. 
	The only one thing left to be proven is that the optimal buyer strategy will either pass trough the pricing of $\T(\tilde{\n})$, or will have a linear regret.
	
	Let $\n_{1}\rightarrow\ldots\rightarrow\n_{\tilde{t}}$ be the path from the root $\n_{1} =\treeroot(\T(\A))$ to the node $\n_{\tilde{t}} = \tilde{\n}$.
	If, for some $t=1,\ldots,\tilde{t}-1$, we have $p^{\n_{t}} =  p^{\n_{t+1}},$ then $p^{\rt(\n_{t})} = p^{\lf(\n_{t})}$, since, otherwise, by the regularity of $\A$ (see Definition~\ref{def_RWC_alg}), we would have: 
	\begin{itemize}
		\item either $\RR(\n_{t})=\RR(\n_{t+1})$ for $p^{\n_{t}} = 0$ ($\LL(\n_{t})=\LL(\n_{t+1})$ for $p^{\n_{t}} = 1$), that contradicts to the definition of $\tilde{\n}$ with the smallest depth; 
		\item or $p^{\m}=p^{\n_{t}}\fa\m\in\T(\n_{t+1})$, that contradicts to the existence of $\tilde{\n}$ with $p^{\tilde{\n}}\in(0,1)$.
	\end{itemize}
	Hence, in this case of $p^{\n_{t}} =  p^{\n_{t+1}},$  by regularity of $\A$,  we have that:
	\begin{itemize}
		\item either the buyer decision at the node $\n_{t}$ does not affect the further pricing: $\RR(\n_{t}) = \LL(\n_{t})$, i.e., the optimal buyer strategy may not pass exactly through the edge $\n_{t}\rightarrow\n_{t+1}$, but, if the buyer select the other edge from the node $\n_{t}$, he  will face the subtree which is price equivalent to the subtree $\T(\n_{t+1})$;
		\item or $\n_{t+1}=\rt(\n_{t})$ for $p^{\n_{t}} = 0$ ($\n_{t+1}=\lf(\n_{t})$ for $p^{\n_{t}} = 1$) and $p^{\m}=p^{\n_{t}}\fa\m\in\LL(\n_{t})$ ($\fa\m\in\RR(\n_{t})$, resp.); thus, if the optimal strategy passes through the alternative node $\lf(\n_{t})$ ($\rt(\n_{t})$, resp.), then the seller will get a linear regret.
	\end{itemize}
	
	If $p^{\n_{t+1}} \neq p^{\n_{t}} = 0$, $t=1,..,\tilde{t}-1$, then, again by the regularity of $\A$, any sub-strategy in the left subtree $\LL(\n_t)$ (a path starting from $\lf(\n_t)$, i.e., from the alternative  to the choice of the right child $\n_{t+1}$ decision) has one of the following forms: 
	\begin{itemize}
		\item (a) there is no any acceptance; 
		\item there is an acceptance and after the first acceptance the buyer either
		\begin{itemize}
			\item (b) will receive pricing of the tree $\RR(\n_t)$; or
			\item (c) will always receive the price $0$.
		\end{itemize}
	\end{itemize}
	If the buyer uses a strategy from the cases~(a) and~(c), then the seller will get a linear regret. The case~(b) means that the algorithm $\A$ will behave similarly whenever the buyer accepts the price $0$: at the round $t^{\n_t}$ or after several rejections. Hence, the strategic buyer will accept $0$ at the round $t^{\n_t}$ (i.e., the buyer follows the edge $\n_{t}\rightarrow\n_{t+1}$).
	The examination of the case $p^{\n_{t+1}} \neq p^{\n_{t}} = 1, t=1,..,\tilde{t}-1$ is similar.
\end{proof}

\subsection{Missed proofs from Section~\ref{subsec_PreTransf}}

\subsubsection{Proof of Lemma~\ref{lemma_pre_RC_alg}}
\label{subsubsec_proof_lemma_pre_RC_alg}
\begin{proof}
	First, for each node $\n\in\T(\pre(q,\A))$, the recursion in Eq.~(\ref{eq_def_pre_transform}) implies that there exists a node $\m\in\T(\A)$ s.t.\ $\LL(\n) \cong \pre\big(p^n,\LL(\m)\big)$ and $\RR(\n) \cong \pre\big(p^{\m},\RR(\m)\big)$. 
	In particular, $p^{\rt(\n)}=p^{\m}$, $p^{\lf(\n)}=p^{\n}$, $\PP(\LL(\n))=\PP(\LL(\m))\cup\{p^{\n}\}$, and  $\PP(\RR(\n))=\PP(\RR(\m))\cup\{p^{\m}\}$. 
	Let us prove by induction that $p^\n\le p \fa p\in\PP(\T(\n))$: (a) this condition (the basis of the induction) is satisfied by the root node $\treeroot(\T(\pre(q,\A)))$ due to the choice of $q$; and (b) the inductive step holds due to $p^{\lf(\n)}=p^{\n}\le p \fa p\in\PP(\LL(\n))\subseteq\PP(\T(\n))$ and $p^{\rt(\n)} = p^{\m}\le p \fa p\in\PP(\RR(\n))=\PP(\RR(\m))\cup\{p^{\m}\}$, where we used $p^{\m}\le p \fa p\in\PP(\RR(\m))$  since the algorithm $\A$ is right-consistent.
	
	Second, note that $p^\n\le p \fa p\in\PP(\T(\n))\supseteq\PP(\RR(\n))$, i.e., the definition of a right-consistent algorithm holds. Therefore,  $\pre(q,\A)\in\mathbf{C_R}$ and the right-side part of weak consistency holds as well.
	Third, $ p^{\lf(\n)} = p^\n \fa\n \in \T(\pre(q,\A))$ as we noted above, and, hence, the left-side part of weak consistency is satisfied (the case of $p^{\lf(\n)} \neq p^\n$ in Definition~\ref{def_WC_alg} of $\mathbf{WC}$ is never realized).
\end{proof}

\subsection{Missed proofs from Section~\ref{subsec_WC_OptAlg}}

\subsubsection{Proof of Proposition~\ref{prop_reject_bound_for_pre}}
\label{subsubsec_proof_prop_reject_bound_for_pre}
\begin{proof}
	Similarly to the proof of Prop.~\ref{prop_reject_bound}, let $S(\m)$ be the surplus obtained by the buyer when playing an
	optimal strategy against $\A$ after reaching the node $\m$, for each node $\m\in\T(\A)$.
	Since the price $p^{\n}$ is rejected then the following inequality holds (see \cite[Lemma~1]{2014-NIPS-Mohri} or the proof of Prop.~\ref{prop_reject_bound})
	\begin{equation}
	\label{prop_reject_bound_for_pre_proof_eq_1}
	\g^{t^\n-1}(v-p^{\n}) + S(\rt(\n)) <  S(\lf^{r}(\n)).
	\end{equation}
	The left subtree's surplus $S(\lf^r(\n))$ can be upper bounded as follows (using $p^{\n}\le p^{\m}  \: \fa\m\in\LL(\lf^{r-1}(\n))$):
	$$
	S(\lf^r(\n)) \le \sum\limits_{t=t^\n+r}^T\g^{t-1}(v - p^{\n}) < \frac{\g^{t^\n+r-1}}{1-\g}(v - p^{\n});
	$$
	while, in contrast to the proof of Prop.~\ref{prop_reject_bound}, we lower bound the right subtree's surplus $S(\rt(\n))$  by $\g^{t^\n}(v-p^{\rt(\n)})$\footnote{This term may be negative (when $v<p^{\rt(\n)}$), but the lower bound on optimal surplus $S(\rt(\n))$ holds a fortiori  in this case.}, because, after accepting $p^{\n}$ at the round $t^{\n}$, the buyer is able to earn at least this amount at the round $t^{\n}+1$.
	We plug these bounds in Eq.~(\ref{prop_reject_bound_for_pre_proof_eq_1}), divide by $\g^{t^\n-1}$, and obtain
	$$
	(v - p^{\rt(\n)} + p^{\rt(\n)} - p^{\n}) + \g(v-p^{\rt(\n)}) < \frac{\g^{r}}{1-\g}(v - p^{\rt(\n)} + p^{\rt(\n)} - p^{\n}) \Leftrightarrow
	$$
	$$
	\Leftrightarrow
	(v - p^{\rt(\n)})\left( 1 + \g - \frac{\g^r}{1-\g}\right)    <  \left(\frac{\g^{r}}{1-\g} - 1\right) (p^{\rt(\n)} - p^{\n}),
	$$
	that implies Eq.~(\ref{prop_reject_bound_for_pre_eq_1}), since $r > \log_\g(1-\g^2)$ implies $1-\g^2-\g^r>0$. 
\end{proof}

\subsubsection{Proof of Proposition~\ref{prop_exploitation_guarantee}}
\label{subsubsec_proof_prop_exploitation_guarantee}
\begin{proof}
	As in the proofs of Prop.~\ref{prop_reject_bound} and Prop.~\ref{prop_reject_bound_for_pre}, let $S(\m)$ be the surplus obtained by the buyer when playing an optimal strategy against $\A$ after reaching the node $\m$, for each node $\m\in\T(\A)$.
	The condition $v < p \fa p\in\PP(\RR(\n))$ implies that $S(\rt(\n)) = 0$ and the strategic buyer will thus gain exactly $\g^{t^\n-1}(v-p^n)$ if he accepts the price $p^n$ at the round $t^\n$. Let us show that there exists a strategy in $\LL(\n)$ with a larger surplus. Indeed, if the buyer rejects $r$ times the price $p^\n$ and accepts this price $G$ times after that, then he gets the following surplus:
	$$
	\sum_{s=t^\n+r}^{t^\n+r+G-1}\g^{s-1}(v-p^n) = \frac{\g^{t^n+r -1} - \g^{t^n+r-1+G}}{1-\g}(v-p^n) = \g^{t^n-1}\g^{r}\frac{1-\g^G}{1-\g}(v-p^n) > \g^{t^\n-1}(v-p^n),
	$$
	where the last inequality holds due to the condition on $G$ and  
	$$
	\g^{r}(1-\g^G)/(1-\g)>1 \Leftrightarrow (1-\g^G)>(1-\g)\g^{-r} \Leftrightarrow \g^G< 1- (1-\g)\g^{-r}.
	$$
\end{proof}

\subsubsection{Proof of Lemma~\ref{lemma_penalty_number_estimate}}
\label{subsubsec_proof_lemma_penalty_number_estimate}
\begin{proof}
	In order to get the claims of this lemma, one just needs to straightforwardly verify few inequalities. Namely, for Prop.~\ref{prop_reject_bound_for_pre}, we have:
	$$
	r = \lceil r_{\g,\varkappa} \rceil \ge r_{\g,\varkappa} = \log_\g\left((1-\g)\Big(1+\frac{\varkappa}{1+\varkappa}\g\Big)\right) > \log_\g\left((1-\g)(1+\g)\right) = \log_\g(1-\g^2)
	$$
	since $\varkappa/(1+\varkappa) < 1 \fa \varkappa > 0$; and
	$$
	\eta_{r,\g} = \frac{\g^r+\g - 1}{1-\g^2-\g^r} 
	\le \frac{(1-\g)\Big(1+\frac{\varkappa}{1+\varkappa}\g\Big) +\g - 1}{1-\g^2-(1-\g)\Big(1+\frac{\varkappa}{1+\varkappa}\g\Big)} 
	= \frac{1+\frac{\varkappa}{1+\varkappa}\g - 1}{1+\g-1-\frac{\varkappa}{1+\varkappa}\g} = \varkappa. 
	$$
	For Prop.~\ref{prop_exploitation_guarantee}, we have:
	$$
	r = \lceil r_{\g,\varkappa} \rceil < r_{\g,\varkappa} + 1 
	= \log_\g(1-\g) + \log_\g\left(\Big(1+\frac{\varkappa}{1+\varkappa}\g\Big)\g\right) <   \log_\g(1-\g)
	$$
	since $\Big(1+\frac{\varkappa}{1+\varkappa}\g\Big)\g>1$ due to $\varkappa>(1-\g)/(\g^2+\g-1)$; and, finally, 
	$$
	G \ge G_{\g,\varkappa}= \log_\g\left(1 - \Big(1+\frac{\varkappa}{1+\varkappa}\g\Big)^{-1}\g^{-1}\right) 
	> \log_\g\left(1 - (1-\g)\g^{-r}\right),
	$$
	where we used $\Big(1+\frac{\varkappa}{1+\varkappa}\g\Big)\g < \g^{r}/(1-\g)$ since $r < r_{\g,\varkappa} + 1$.
\end{proof}

\subsubsection{Proof of Theorem~\ref{th_prePRRFES_regret_upper_bound}}
\label{subsubsec_proof_th_prePRRFES_regret_upper_bound}
\begin{proof}
	Note that the conditions of this theorem allow us to apply Lemma~\ref{lemma_penalty_number_estimate}, Prop.~\ref{prop_reject_bound_for_pre}, and Prop.~\ref{prop_exploitation_guarantee}, that make the other technique of the proof similar to the one of Theorem~\ref{th_PRRFES_regret_upper_bound}.
	So, let $L$ be the number of phases conducted by the algorithm during $T$ rounds, then we decompose the total regret over $T$ rounds into the sum of the phases' regrets: $\SReg(T,\A, v, \gb) = \sum_{l=0}^{L}R_l$. For the regret $R_l$ at each phase except the last one, the following identity holds:
	\begin{equation}
	\label{th_prePRRFES_regret_upper_bound_proof_eq1}
	R_l =  \sum\limits_{k=0}^{K_l-1} (v-p_{l,k}) + rv + g(l)(v-p_{l,K_{l}}), \quad l=0,\ldots,L-1,
	\end{equation} 
	where the first, second, and third terms correspond to the exploration rounds with acceptance, the reject-penalization rounds, and the exploitation rounds, respectively.
	First, note that here, in the exploitation rounds, we directly use Proposition~\ref{prop_exploitation_guarantee} (via Lemma~\ref{lemma_penalty_number_estimate} since $g(l)\ge G_{\g,\varkappa}$) to conclude that $p_{l, K_{l}}<v$ and the price $p_{l, K_{l}}$ is thus accepted by the strategic buyer at the exploitation rounds (since the buyer's decisions at these rounds do not affect further pricing of the algorithm prePRRFES and $p_{l, K_{l}}<v$).

	Second, since the price $p_{l,K_{l}}$ is rejected, we have $v - p_{l,K_{l} + 1} < \varkappa(p_{l,K_{l} + 1} - p_{l,K_{l}})=\varkappa\e_l$ (by Proposition~\ref{prop_reject_bound_for_pre} via Lemma~\ref{lemma_penalty_number_estimate} since $\eta_{r,\g}\le\varkappa$ for $r\ge \lceil r_{\g,\varkappa} \rceil$ and any $t\in\mathbb{N}$). Hence, the valuation 
	$v\in\big(p_{l,K_{l}}, p_{l,K_{l}} + (1+\varkappa)\e_{l}\big)$ and all accepted prices $p_{l+1,k}, \fa k\le K_{l+1}$, from the next phase $l+1$ satisfy: 
	$$
	p_{l+1,k}\in(q_{l+1},v)\subseteq \big(p_{l,K_{l}}, p_{l,K_{l}} + (1+\varkappa)\e_{l}\big) \quad \fa k\le K_{l+1},
	$$ 
	because any accepted price has to be lower than the valuation~$v$ for the strategic buyer (whose optimal strategy is locally non-losing one for $\A\in\mathbf{C_R}$, see the discussion after \cite[Lemma~1]{2017-WWW-Drutsa}). This infers $K_{l+1} <(1+\varkappa)N_{l+1}\le\big\lceil(1+\varkappa)N_{l+1}\big\rceil=:N_{l+1,\varkappa}$ since $N_{l+1} = \e_l/\e_{l+1}$ by Eq.~(\ref{eq_prePRRFES_def_round_parameters}). Therefore, for the phases $l=1,\ldots,L$, we have:
	\begin{equation*}
	v - p_{l,K_{l}} < (1+\varkappa)\e_{l}; \qquad
	v-p_{l,k} < \e_{l}\big((1+\varkappa)N_{l}-k\big)  \fa k\in\mathbb{Z}_{N_{l,\varkappa}};
	\end{equation*}
	and
	\begin{equation*}
	\begin{split}
	\sum\limits_{k=0}^{K_{l}-1}(v-p_{l,k}) < \e_{l}\sum\limits_{k=0}^{N_{l,\varkappa}-2}\big((1+\varkappa)N_{l}-k\big) = \e_{l}\frac{N_{l,\varkappa}-1}{2}\big(2(1+\varkappa)N_l-N_{l,\varkappa}+2\big) \le
	\\
	\le \e_{l}\frac{(1+\varkappa)N_l}{2}\big((1+\varkappa)N_l+2\big) = \frac{(1+\varkappa)^2}{2}N_l\cdot N_l\e_{l} + (1+\varkappa)N_l\e_l =  \frac{(1+\varkappa)^2}{2} + (1+\varkappa)\e_{l-1},
	\end{split}
	\end{equation*} 
	where we used the definitions of $N_l$ and $\e_l$ (i.e., $N_l\e_l = \e_{l-1}$ and $N_l = \e_l^{-1}$), as in the proof of Theorem~\ref{th_PRRFES_regret_upper_bound}.
	For the zeroth phase $l=0$, one has trivial bound 
	$\sum_{k=0}^{K_{0}-1}(v-p_{0,k}) \le 1$.
	Hence, by definition of the exploitation rate $g(l)$, we have 
	$$g(l)\cdot\e_l=\max\{\e_l^{-1}\cdot\e_l,\lceil G_{\g,\varkappa}\rceil\cdot\e_l\} \le \max\{1,\lceil G_{\g,\varkappa}\rceil/2\},$$
	and, thus,
	\begin{equation}
	\label{th_prePRRFES_regret_upper_bound_proof_eq2} 
	R_l \le  \frac{(1+\varkappa)(2+\varkappa)}{2} + rv + g(l) \cdot (1+\varkappa)\e_{l} \le  rv+\frac{(1+\varkappa)}{2}(2+\max\{2,\lceil G_{\g,\varkappa}\rceil\}+\varkappa), \quad l=0,\ldots,L-1.
	\end{equation} 
	
	The $L$-th phase differs from the other ones only in possible absence of some rounds: (reject-penalization or exploitation ones). In this phase, we consider two cases on the actual number of exploitation rounds $g_L(L)$:  (a) $g_L(L)\ge \lceil G_{\g,\varkappa}\rceil$ and (b) $g_L(L) < \lceil G_{\g,\varkappa}\rceil$. In the case~(a), we again apply Proposition~\ref{prop_exploitation_guarantee} (via Lemma~\ref{lemma_penalty_number_estimate} since $g_L(L)\ge \lceil G_{\g,\varkappa}\rceil$) to get that $p_{L, K_{L}}<v$ and the price $p_{L, K_{L}}$ is thus accepted by the strategic buyer at the exploitation rounds. In this case, we have thus:
	\begin{equation}
	\label{th_prePRRFES_regret_upper_bound_proof_eq3} 
	R_L =  \sum\limits_{k=0}^{K_L-1} (v-p_{L,k}) + rv + g_L(L)(v-p_{L,K_{L}}).
	\end{equation}
	The right-hand side of Eq.~(\ref{th_prePRRFES_regret_upper_bound_proof_eq3}) is upper-bounded by the right-hand side of Eq.~(\ref{th_prePRRFES_regret_upper_bound_proof_eq1}) with $l=L$, which is in turn upper-bounded by the right-hand side of Eq.~(\ref{th_prePRRFES_regret_upper_bound_proof_eq2}).
	In the case~(b), we have no guarantee that $p_{L, K_{L}}<v$ and, hence, $p_{L, K_{L}}$ may be rejected by the strategic buyer at the exploitation rounds. Hence, we have to estimate the regret in the last phase in the following way:
	\begin{equation}
	\label{th_prePRRFES_regret_upper_bound_proof_eq4} 
	R_L =  \sum\limits_{k=0}^{K_L-1} (v-p_{L,k}) + r_Lv + g_L(L)v \le \frac{(1+\varkappa)(2+\varkappa)}{2} + (r + \lceil G_{\g,\varkappa}\rceil -1)v,
	\end{equation}
	where $r_L$ the actual number of reject-penalization rounds, $r_L\le r$.
	
	Finally, using $(\lceil G_{\g,\varkappa}\rceil -1)v-\max\{1,\lceil G_{\g,\varkappa}\rceil/2\} \le \lceil G_{\g,\varkappa}\rceil/2 - 1$, one has
	\begin{equation*}
	\SReg(T,\A, v, \gb) = \sum_{l=0}^{L}R_l \le \left(rv+\frac{(1+\varkappa)}{2}(2+\max\{2,\lceil G_{\g,\varkappa}\rceil\}+\varkappa)\right)(L+1) + \frac{\lceil G_{\g,\varkappa}\rceil}{2} - 1.
	\end{equation*} 
	Thus, one needs only to estimate the number of phases $L$ by the number of rounds $T$. So, for $2\le T\le 2 + r + g(0)$, we have $L=0$ or $1$ and thus $L+1\le 2 \le \log_2\log_2 T + 2$. For $T \ge 2 + r + g(0)$, we have $T = \sum_{l=0}^{L-1}(K_l + r + g(l)) + K_L + r_L + g_L(L)\ge g(L-1)$ with $L>0$.
	Hence, $2^{2^{L-1}} \le g(L-1) \le  T$, which implies $L \le \log_2\log_2 T + 1$. Summarizing, we get Eq.~(\ref{th_prePRRFES_regret_upper_bound_eq1}).
\end{proof}

\begin{figure}
	\centering
	\includegraphics[width=\textwidth]{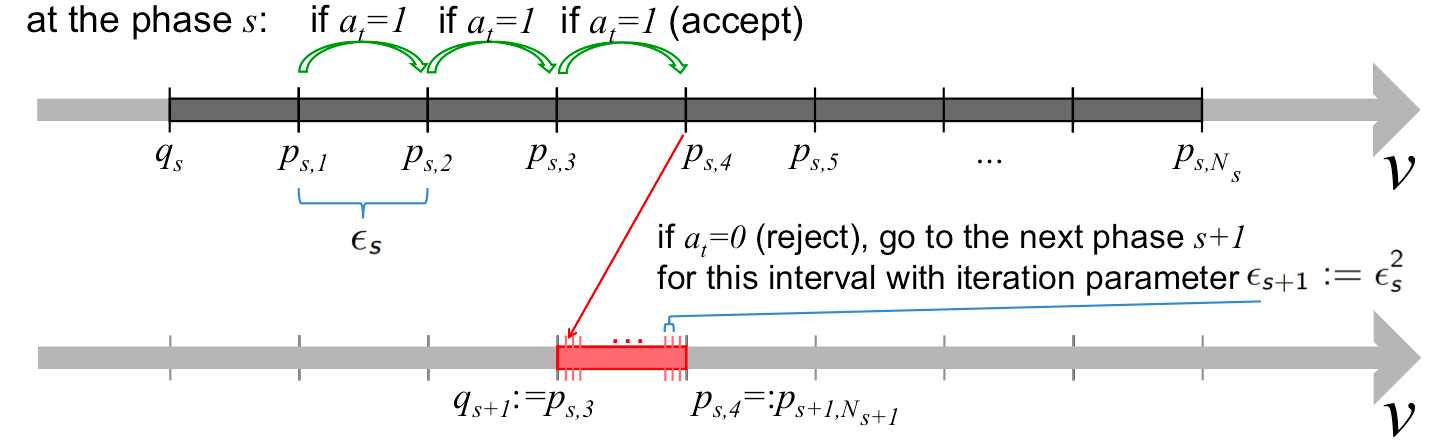}
	\caption{The work flow of the Fast Search step: the prices offered at the phases $s$ and $s+1$.}
	\label{img_FS_alg}
\end{figure}

\section{Penalized Reject-Revising Fast Exploiting Search (PRRFES)}
\label{sec_PRRFES_descr}

The pricing algorithm \emph{Penalized Reject-Revising Fast Exploiting Search} (\emph{PRRFES}) was presented in~\cite{2017-WWW-Drutsa}. 
It is a special improvement of the pricing algorithm Fast Exploiting Search (\emph{FES}), which   was presented in~\cite{2017-WWW-Drutsa} as well and is  in turn a horizon-independent improvement of the pricing algorithm Fast Search~\cite{2003-FOCS-Kleinberg} designed to act against a myopic (truthful) buyer with a tight regret bound in $\Theta(\log\log T)$.
The key peculiarities of PRRFES consist in utilization of penalization rounds after a rejection, forcing thus the buyer to lie less (similarly to~\cite{2014-NIPS-Mohri}), andß in a regular revising of rejected prices.

Namely, the pricing algorithm PRRFES
works in phases initialized by the phase index $l:=0$, the last accepted price before the current phase $q_0:=0$, the iteration parameter $\e_0:=1/2$, and the number of offers $N_0:=2$; at each phase $l\in\mathbb{Z}_{+}$, it sequentially offers prices $p_{l,k}:= q_{l} + k\e_{l}, k\in\mathbb{N}$ (i.e., in contrast to FES, $k$ can now be  higher than $N_l$, thus, it can \underline{explore} prices higher than the earlier rejected one $p_{l,N_l}=p_{l-1,K_{l-1}+1}$), with $\e_{l}$ and $N_l$ defined in Eq.~(\ref{eq_FES_def_round_parameters}):
\begin{equation}
\label{eq_FES_def_round_parameters}
\e_l := \e^2_{l-1} = 2^{-2^l}, 
\: N_l := \e_{l-1}/\e_{l} = \e^{-1}_{l-1} = 2^{2^{l-1}}, 
\: l\in\mathbb{N};
\end{equation}
if a price $p_{l,k}$ with $k = K_{l}+1\ge 1$ is rejected,  (1) it offers this price $p_{l,K_{l}+1}$ for $r-1$ rounds (\underline{penalization}: if one of them is accepted, PRRFES continues offering $p_{l,k}, k=K_{l}+2,..$ following the Definition~\ref{def_PenalNodeSeq}), (2) it offers the price $p_{l,K_{l}}$ for $g(l)$ rounds   (\underline{exploitation}), and (3) PRRFES goes to the next phase by setting $q_{l+1}:=p_{l,K_{l}}$ and $l:=l+1$.
The pseudo-code of PRRFES is presented in Alg.~\ref{alg_PRRFES}, which is in the class $\mathbf{C_R}$ (i.e., it is right-consistent) and does not belong to the class $\mathbf{WC}$ (i.e., it is not weakly consistent). 
In Fig.~\ref{img_FS_alg}, we present a scheme of the work flow of the Fast Search step.

\section{Pseudo-codes of algorithms}
We present pseudo-codes for the following algorithms:
\begin{itemize}
	\item a general horizon-independent algorithm $\A$ in Algorithm~\ref{alg_A}
	\item $\pre(q,\A)$, a $\pre$-transformation of a general horizon-independent algorithm $\A$ in Algorithm~\ref{alg_preA};
	\item PRRFES in Algorithm~\ref{alg_PRRFES};
	\item prePRRFES in Algorithm~\ref{alg_prePRRFES}.
\end{itemize}

\begin{figure*}[ttt!]
	\begin{minipage}[t]{0.5\textwidth}
		\vspace{0pt} 
		\begin{algorithm}[H]
			\small
			\caption{Pseudo-code of $\A$.}
			\label{alg_A}
			\begin{algorithmic}[1]
				\STATE {{\bfseries Input:}  $\A\in\ASet$}
				\STATE {{\bfseries Initialize:} $\n := \treeroot(\T(\A)),$}
				\WHILE{the buyer plays}
				\STATE {Offer the price $p^\n$ to  the buyer}
				\IF{the buyer accepts the price}
				\STATE {$\n := \rt(\n)$}
				\ELSE
				\STATE {$\n := \lf(\n)$}
				\ENDIF
				\ENDWHILE		
			\end{algorithmic}
		\end{algorithm}
	\end{minipage}
	\begin{minipage}[t]{0.5\textwidth}
		\vspace{0pt} 
		\begin{algorithm}[H]
			\small
			\caption{Pseudo-code of $\pre(q,\A)$.}
			\label{alg_preA}
			\begin{algorithmic}[1]
				\STATE {{\bfseries Input:}  $q\in\mathbb{R}$ and $\A\in\ASet$}
				\STATE {{\bfseries Initialize:} $\n := \treeroot(\T(\A)),\: p := q$}
				\WHILE{the buyer plays}
				\STATE {Offer the price $p$ to  the buyer}
				\IF{the buyer accepts the price}
				\STATE {$p := p^\n$}
				\STATE {$\n := \rt(\n)$}
				\ELSE
				\STATE {$\n := \lf(\n)$}
				\ENDIF
				\ENDWHILE		
			\end{algorithmic}
		\end{algorithm}
	\end{minipage}
	\vspace{-1em}
\end{figure*}

\begin{figure*}[ttt!]
	\begin{minipage}[t]{0.5\textwidth}
		\vspace{0pt} 
		\begin{algorithm}[H]
	\small
	\caption{Pseudo-code of the {\small PRRFES~\cite{2017-WWW-Drutsa}}}
	\label{alg_PRRFES}
	\begin{algorithmic}[1]
		\STATE {{\bfseries Input:} $r\in\mathbb{N}$ and $g:\mathbb{Z}_{+}\rightarrow\mathbb{Z}_{+}$} 		
		\STATE {{\bfseries Initialize:} $q := 0,\: p := 1/2,\:  l := 0$}
		\WHILE{the buyer plays}
		\STATE {Offer the price $p$ to  the buyer}
		\IF{the buyer accepts the price}
		\STATE {$q := p$ \label{alg_PRRFES_line_accept}}
		\ELSE
		{
			\STATE {Offer the price $p$ to the buyer for $r-1$ rounds\!\!\!}
			\IF{the buyer accepts one of the prices}
			\STATE {{\bf go to} line~\ref{alg_PRRFES_line_accept}} 
			\ENDIF}
		\STATE {Offer the price $q$ to the buyer for $g(l)$ rounds\!\!} 
		\STATE {$l := l + 1$}
		\ENDIF
		{
			\IF{$p < 1$}
			\STATE {$p := q + 2^{-2^l}$}
			\ENDIF}
		\ENDWHILE		
	\end{algorithmic}
\end{algorithm}
	\end{minipage}
	\begin{minipage}[t]{0.5\textwidth}
		\vspace{0pt} 
		\begin{algorithm}[H]
	\small
	\caption{Pseudo-code of the {\small prePRRFES}}
	\label{alg_prePRRFES}
	\begin{algorithmic}[1]
		\STATE {{\bfseries Input:} $r\in\mathbb{N}$ and $g:\mathbb{Z}_{+}\rightarrow\mathbb{Z}_{+}$} 		
		\STATE {{\bfseries Initialize:} $q := 0,\: p := 1/2,\:  l := 0$}
		\WHILE{the buyer plays}
		{\color{blue}\STATE  Offer the price $q$ to  the buyer}
		\IF{the buyer accepts the price}
		\STATE {$q := p$ \label{alg_prePRRFES_line_accept}}
		\ELSE
		{
			{\color{blue}\STATE {Offer the price $q$ to the buyer for $r-1$ rounds\!\!\!}}
			\IF{the buyer accepts one of the prices}
			\STATE {{\bf go to} line~\ref{alg_prePRRFES_line_accept}} 
			\ENDIF}
		\STATE {Offer the price $q$ to the buyer for $g(l)$ rounds\!\!} 
		\STATE {$l := l + 1$}
		\ENDIF
		{
			\IF{$p < 1$}
			\STATE {$p := q + 2^{-2^l}$}
			\ENDIF}
		\ENDWHILE		
	\end{algorithmic}
\end{algorithm}

	\end{minipage}
	\vspace{-1em}
\end{figure*}

\bibliographystyle{abbrv}
\bibliography{2017-arxiv-hipra}

\medskip

\end{document}